\documentclass[aps,prl,twocolumn,superscriptaddress]{revtex4-1}
\usepackage{graphicx,amssymb,amsmath,amsfonts,empheq}
\usepackage[dvipsnames]{xcolor}
\usepackage{color}
\usepackage{braket}
\usepackage{latexsym}
\usepackage{upgreek}
\usepackage{dsfont}
\usepackage{bm}
\usepackage{multirow}
\usepackage{enumitem}
\usepackage[normalem]{ulem}
\usepackage{url}
\usepackage{hyperref}
\hypersetup{
colorlinks = true,
linkcolor = [rgb]{0.70,0.13,0.13},
citecolor = [rgb]{0.13,0.55,0.13},
urlcolor = [rgb]{0.25, 0.41, 0.88}}
\usepackage{bbm}

\newenvironment{rmk}[1][Remark]{\begin{trivlist}
\item[\hskip \labelsep {\bfseries #1}]}{\end{trivlist}}

\usepackage{algorithm}
\usepackage[]{algpseudocode}

\makeatletter
\newenvironment{breakablealgorithm}
  {
   \begin{center}
     \refstepcounter{algorithm}
     \hrule height.8pt depth0pt \kern2pt
     \renewcommand{\caption}[2][\relax]{
       {\raggedright\textbf{\ALG@name~\thealgorithm} ##2\par}%
       \ifx\relax##1\relax 
         \addcontentsline{loa}{algorithm}{\protect\numberline{\thealgorithm}##2}%
       \else 
         \addcontentsline{loa}{algorithm}{\protect\numberline{\thealgorithm}##1}%
       \fi
       \kern2pt\hrule\kern2pt
     }
  }{
     \kern2pt\hrule\relax
   \end{center}
  }
\makeatother

\renewcommand{\geq}{\geqslant}
\renewcommand{\Re}{\operatorname{Re}}
\renewcommand{\Im}{\operatorname{Im}}

\newcommand{\Tr}{\operatorname{Tr}}

\newcommand{\U}{\operatorname{U}}

\newcommand{\T}{\operatorname{\textbf{T}}}

\newcommand{\ii}{\mathrm{i}}
\newcommand{\dd}{\mathrm{d}}

\newcommand{\II}{\mathbb{I}}
\newcommand{\id}{\mathbb{I}}

\newcommand{\ZZ}{\mathbb{Z}}
\newcommand{\EE}{\mathbb{E}}

\newcommand{\scE}{\mathcal{E}}

\newcommand{\scO}{\mathcal{O}}

\newcommand{\scX}{\mathcal{X}}

\newcommand{\eqnref}[1]{Eq.~\eqref{#1}}
\newcommand{\figref}[1]{Fig.~\ref{#1}}
\newcommand{\appref}[1]{Appendix~\ref{#1}}

\newcommand{\refcite}[1]{Ref.\,\onlinecite{#1}}

\newcommand{\norm}[1]{{\lVert #1\rVert}}
\newcommand{\mat}[1]{\left[\begin{matrix}#1\end{matrix}\right]}

\newcommand{\dia}[3]{\raisebox{#3pt}{\includegraphics[height=#2pt]{dia_#1}}}
\newcommand{\eq}[1]{\begin{equation}#1\end{equation}}
\newcommand{\eqs}[1]{\begin{equation}\begin{split}#1\end{split}\end{equation}}

\begin{document}

\title{Self-Organized Error Correction in Random Unitary Circuits with Measurement}

\author{Ruihua Fan}
\affiliation{Department of Physics, Harvard University, Cambridge, MA 02138, USA}
\author{Sagar Vijay}
\affiliation{Department of Physics, Harvard University, Cambridge, MA 02138, USA}
\author{Ashvin Vishwanath}
\affiliation{Department of Physics, Harvard University, Cambridge, MA 02138, USA}
\author{Yi-Zhuang You}
\affiliation{Department of Physics, University of California, San Diego, La Jolla, CA 92093, USA}

\begin{abstract}

Random measurements have been shown to induce a phase transition in an extended quantum system evolving under chaotic unitary dynamics, when the strength of measurements exceeds a threshold value.  Below this threshold, a steady state with a sub-thermal volume law  entanglement emerges, which is resistant to the disentangling action of measurements, suggesting a connection to quantum error-correcting codes.   Here we quantify these notions by identifying a universal, subleading logarithmic contribution to the volume law entanglement entropy: $S^{(2)}(A)=\kappa L_A+\frac{3}{2}\log L_A$ which bounds the mutual information between a qudit inside region $A$ and the rest of the system. Specifically, we find the power law decay of the mutual information $I(\{x\}:\bar{A})\propto x^{-3/2}$   with distance $x$ from the region's boundary, which implies that measuring a qudit deep inside $A$  will have negligible effect on the entanglement of $A$. We obtain these results by mapping the entanglement dynamics to the imaginary time evolution of an Ising model, to which we can apply field-theoretic and matrix-product-state techniques.  Finally, exploiting the error-correction viewpoint, we assume that the volume-law state is an encoding of a Page state in a quantum error-correcting code to obtain a bound  on the critical measurement strength $p_{c}$ as a function of the qudit dimension $d$: $p_{c}\log[(d^{2}-1)({p_{c}^{-1}-1})]\le \log[(1-p_{c})d]$. The bound is saturated at $p_c(d\rightarrow\infty)=1/2$ and provides a reasonable estimate for the qubit transition: $p_c(d=2) \le 0.1893$.

\end{abstract}
\maketitle


\emph{Introduction}---
The study of random unitary circuits has significantly advanced our understanding of the universal behavior of entanglement and operator dynamics in quantum many-body systems\cite{Nahum2017Quantum,Zhou2018Emergent,Keyserlingk2018Operator,Nahum2018Operator,Nahum2018Dynamics,Vijay2018Finite-Temperature,You2018Entanglement,Rakovszky2018Diffusive,Chan2018Spectral,Chan2018Solution,Bertini2019ESMMMMQC,Rakovszky2019Entanglement}. Locally accessible quantum information becomes   scrambled under unitary evolution\cite{Lashkari2013TFSC,Hosur2016Chaos}, which typically leads to thermalization, accompanied by volume-law scaling of the entanglement entropy\cite{Page:1993fv} in the steady-state.  This is consistent with the idea that entropy should be an extensive property for thermal systems\cite{Deutsch1991Quantum,Srednicki1994Chaos}. It has been recently found that performing local measurements along with random, local  unitary dynamics can slow down and stop thermalization. Conditioned on the measurement outcome, the qubit that has been measured will be projected to a product state, and disentangled from the rest of the system. When the measurement rate is high enough, most qubits in the system will be disentangled, and the final state will exhibit area-law entanglement scaling\cite{Srednicki1993Entropy,Verstraete2006Criticality,Hastings2008An-area}, a manifestation of the quantum Zeno effect\cite{Li2018QZEMET}. Driven by the measurement rate, the final state of this quantum channel (i.e.~the quantum circuit with measurements) exhibits an entanglement phase transition driven by the measurement rate \cite{Li2018QZEMET,Chan2019UED,Skinner2019MPTDE}, which has attracted much recent interest\cite{Bao2019TPTRUCWM,Jian2019MCRQC,Li2019METHQC,Choi2019QECEPTRUCWPM,Gullans2019Dynamical,Zabalo2019Critical,Gullans2019Probes,Szyniszewski2019ETFVWM,Tang2019MPTCSNMDRGC}.

Progress has been made in understanding this transition by mapping the problem to the statistical mechanics model of permutation group elements\cite{Vasseur2018Entanglement,Skinner2019MPTDE,Bao2019TPTRUCWM,Jian2019MCRQC}, where the entanglement transition between the volume-law and area-law phases corresponds to the ordering  transition in this classical lattice ``magnet". The universality of the entanglement transition remains to be fully understood, due to the difficulty in taking the required replica limit\cite{Bao2019TPTRUCWM,Jian2019MCRQC} of the statistical mechanics model.

 In this work, we turn our attention away from the transition point to focus on features of the volume-law phase. Specifically, what are the key, quantitative properties of the volume-law phase that ensure its stability against local measurements? To answer this question, we propose a ``mean-field'' description for the measurement-induced entanglement transition based on the recently developed entanglement feature formulation for locally scrambled quantum dynamics\cite{Kuo2019Markovian}, which is in line with the statistical mechanics description of quantum many-body entanglement\cite{Hayden2016Holographic,Jonay2018CDOSE,Mezei2018Membrane}. This mean-field description cannot provide a precise description of the critical fluctuations  at the transition point; nevertheless, it accurately describes the asymptotic entanglement properties away from the transition. 
{The entanglement feature formalism provides a powerful tool for studying unitary dynamics with measurements. Within this formulation, the mean-field description of the evolution of the second R\'{e}nyi entropy is Markovian, as it only relies on the second R\'{e}nyi entropies for all sub-systems in the previous timestep of the evolution.  These entanglement dynamics can be further related to the Floquet dynamics of an Ising model, whose steady-state properties  can be determined by well-developed theoretical and numerical methods}.

Using this solution, we show that the volume-law phase exhibits a universal sub-leading logarithmic entropy scaling. Furthermore, the reduction of the entanglement entropy after performing a measurement decays as a power-law in the distance of the measurement from the region's boundary.  These features suggest the quantum error correction (QEC) property in volume-law states, which accounts for its stability against local measurements. By quantitatively studying the error-correcting properties of the final state, we also derive a bound (\ref{eq:pc_bound}) on the measurement rate as a function of the local Hilbert space dimension, above which the system must be in an area-law entangled phase.  In previous work \cite{Choi2019QECEPTRUCWPM,Gullans2019Dynamical}, other perspectives that relate the entanglement phase transition to QEC have been discussed. 

\begin{figure}[htbp]
	\centering
	\includegraphics[width=0.56\columnwidth]{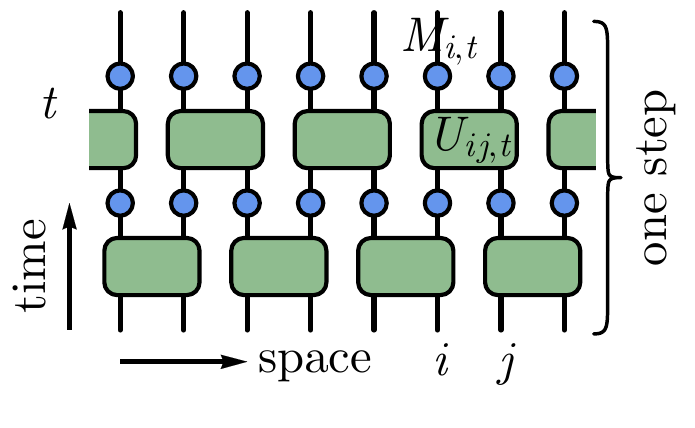}
	\caption{The random quantum channel model. Green blocks are two-qudit Haar random unitary gates. Blue dots are single-qudit random measurements (which can be either weak measurements or projective measurements with probability).}
	\label{fig:circuit}
\end{figure}

\emph{Random Quantum Channel Model}---
We consider the quantum dynamics of a 1D array of $N$ qudits, each with Hilbert space dimension $d$. The quantum dynamics is modeled by a random unitary circuit with random measurements implemented uniformly, as shown in \figref{fig:circuit}. The circuit consists of two-qudit unitary gates $U_{ij,t}$ (acting on qudits $i,j$ at layer $t$) arranged in a brick-wall pattern. All gates are drawn from the Haar random unitary ensemble independently throughout space and time. After each layer of the unitary gates, measurements are carried out on every qudit. Each single-qudit measurement can be described by the measurement operator $M_{i,t}$ (acting on qudit $i$ at layer $t$)\cite{Kraus1983States,Brun2002Simple}, which is independently drawn from the ensemble $\{\id\}\cup\{\sqrt{d}P_V|V\in \U(d)\}$ with the probability measure $P(\id)=1-p$ and $P(\sqrt{d}P_V)=p\,\dd V$ (with $\dd V$ being the Haar measure)\cite{Jian2019MCRQC}, where $P_V=V\ket{0}\bra{0}V^\dagger$ represents a random projector in the qudit Hilbert space. This ensemble can model either a projective measurement\cite{Li2018QZEMET,Skinner2019MPTDE} happening with probability $p$ or a weak measurement\cite{Bao2019TPTRUCWM} with strength $p$. Both the unitary operator $U_{ij,t}$ and the measurement operator $M_{i,t}$ can be generally denoted as the Kraus operator $K_{x,t}$ at different spacetime positions labeled by $(x,t)$ in general. They together form the quantum channel, described by the overall Kraus operator $K=\prod_{t}\prod_{x}K_{x,t}$, such that the density matrix $\rho$ of the quantum system evolves by the completely positive trace-preserving map $\rho\to K\rho K^\dagger/\Tr(K\rho K^\dagger)$ under the quantum dynamics.\footnote{Strictly speaking, our protocol differs from projective  measurement. The latter requires post-selection based on the probability of possible outcomes, while our protocol is restricted to applying projectors. In the volume-law phase, local scrambling and the low-density of measurements implies that the probability of each outcome is identical and the two protocols should yield similar results. For Clifford unitary dynamics with measurements in the Pauli basis, the entanglement properties of the state are independent of the measurement outcomes, so in this case, our protocol is actually identical to performing projective measurements.}

\emph{Entanglement Feature Formalism}---
We are interested in the purity of the reduced density matrix $\rho_A=\Tr_{\bar{A}}\rho$ over all possible subsystem $A$,
\begin{equation}
    W_\rho(A)=\Tr\rho_A^2=e^{-S_{\rho}^{(2)}(A)}.
\end{equation}
It is directly related to the 2nd R\'enyi entanglement entropy $S_{\rho}^{(2)}(A)$ that quantifies the amount of quantum entanglement between $A$ and $\bar{A}$ in the state $\rho$ (assuming $\rho$ is pure). To organize this purity data in a more concise way, we introduce a set of Ising variables $[\sigma]\equiv[\sigma_1,\sigma_2,\cdots,\sigma_N]$ to label the subsystem $A$, s.t.~$\sigma_i=-1\;(\downarrow)$ if $i\in A$ and $\sigma_i=+1\;(\uparrow)$ if $i\in\bar{A}$. Then $W_\rho(A)$ can be written as\cite{You2018Entanglement,Kuo2019Markovian}
\begin{equation}\label{eq:Wrho}
    W_\rho[\sigma]=\Tr\rho^{\otimes 2}\scX_{\sigma},
\end{equation}
where $\scX_\sigma=\prod_i\scX_{\sigma_i}$ is a string of identity $\scX_{\sigma_i=\uparrow}\equiv\dia{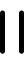}{12}{-3}$ and swap $\scX_{\sigma_i=\downarrow}\equiv\dia{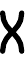}{12}{-3}$ operators acting in the duplicated Hilbert space as specified by the Ising variable $\sigma_i$. The collection of $W_\rho[\sigma]$ over all Ising configurations $[\sigma]$ is called the \emph{entanglement feature}\cite{You2018Machine,You2018Entanglement} of the density matrix $\rho$, which characterizes the entanglement properties of $\rho$. They can be naturally assembled into a vector
\begin{equation}
    \ket{W_\rho}=\sum_{[\sigma]}W_\rho[\sigma]\ket{\sigma},
\end{equation}
called the \emph{entanglement feature state}\cite{Kuo2019Markovian}, with $\ket{\sigma}$ being a set of orthonormal Ising basis labelled by $[\sigma]$. The normalization $\Tr\rho=1$ implies $\braket{\Uparrow|W_\rho}=1$ for the entanglement feature state, where $\ket{\Uparrow}$ denotes the all-up state ($\forall i:\sigma_i=+1$). Nevertheless, $\ket{W_\rho}$ is still well-defined for unnormalized density matrix $\rho$ following \eqnref{eq:Wrho}, which will also be useful in our discussion.

As the state $\rho$ evolves under the random quantum channel in \figref{fig:circuit}, the corresponding entanglement feature state $\ket{W_\rho}$ also evolves, which defines the entanglement dynamics. \refcite{Kuo2019Markovian} pointed out that if the quantum dynamics is locally scrambled, the corresponding entanglement dynamics is Markovian and admits a transfer matrix description. More precisely, suppose the Kraus operator $K$ is randomly drawn from a local-basis independent ensemble, i.e.~the probability $P(K)=P(VKV^\dagger)$ is invariant under arbitrary local (on-site) basis transformation $V=\prod_{i}V_i$ for $V_i\in\U(d)$, then under the completely positive map $\rho_0\to \rho=K\rho_0 K^\dagger$, the corresponding (ensemble averaged) entanglement feature state evolves as
\begin{equation}\label{eq:W=TW}
    \ket{W_{\rho}}\equiv\mathop{\EE}_K\ket{W_{K\rho_0 K^\dagger}}=\hat{W}_K\hat{W}_\id^{-1}\ket{W_{\rho_0}}=\hat{T}_K\ket{W_{\rho_0}},
\end{equation}
where the \emph{entanglement feature operator} $\hat{W}_K$ associated with a Kraus operator $K$ is defined as\cite{You2018Entanglement,Kuo2019Markovian}
\begin{equation}\label{eq:WK}
    \begin{split}
        \hat{W}_K&=\sum_{[\sigma,\tau]}\ket{\sigma}W_K[\sigma,\tau]\bra{\tau},\\
        W_{K}[\sigma,\tau]&=\Tr K^{\dagger\otimes2}\scX_\sigma K^{\otimes2}\scX_\tau,
    \end{split}
\end{equation}
which captures the entanglement feature of the quantum channel $K$ among its input and output degrees of freedoms. Here, $\hat{W}_\id$ is the entanglement feature operator for the identity channel, whose inverse  is denoted by $\hat{W}_\id^{-1}$. The entanglement dynamics is then determined by the transfer matrix $\hat{T}_K=\hat{W}_K\hat{W}_\id^{-1}$, which solely depends on the entanglement property of the quantum channel $K$. 

\emph{Mean-Field Description}---
The random quantum channel model falls in the class of locally scrambled quantum dynamics, for which \eqnref{eq:W=TW} applies. However, \eqnref{eq:W=TW} only provides the average entanglement feature for the unnormalized state $\rho= K\rho_0 K^{\dagger}$. For the normalized final state $\bar{\rho}=\rho/\Tr\rho$, its average entanglement feature 
\begin{equation}\label{eq:avgW1}
    W_{\bar{\rho}}[\sigma]=\mathop{\EE}_{K}\Tr\bar{\rho}^{\otimes2}\scX_\sigma =\mathop{\EE}_{K}\frac{\Tr \rho^{\otimes2}\scX_\sigma}{(\Tr \rho)^2},
\end{equation}
 is still difficult to evaluate. Rigorous treatments have been developed using the replica trick\cite{Vasseur2018Entanglement,Skinner2019MPTDE,Bao2019TPTRUCWM,Jian2019MCRQC,Lopez-Piqueres2020Mean-field}. Nevertheless, we will approximate the average of ratio in \eqnref{eq:avgW1} by the ratio of averages to achieve a simplified ``mean-field'' description
\begin{equation}\label{eq:avgW2}
    W_{\bar{\rho}}[\sigma]\simeq\frac{\mathop{\EE}_{K}\Tr\rho^{\otimes2}\scX_\sigma}{\mathop{\EE}_{K}\Tr \rho^{\otimes2}}=\frac{W_{\rho}[\sigma]}{W_{\rho}[\Uparrow]}=\frac{\braket{\sigma|W_{\rho}}}{\braket{\Uparrow|W_{\rho}}}.
\end{equation}
In this mean-field treatment, we replace the denominator $(\Tr K\rho K^\dagger)^2$ by its expectation value and neglect its fluctuation with respect to $K$. The reason is that the random unitary gates are fast local scramblers, on-site thermalization should be quickly achieved after every layer of unitaries. So the reduced density matrix for each single qudit should look maximally mixed $\rho_i\simeq \id/d$ before the measurement. Then the trace $\Tr M_i\rho_iM_i^\dagger\simeq 1$ is almost independent of the choice of the measurement operator $M_i$ (see \appref{app:check} for a numerical verification), hence the denominator fluctuation should be small. {Although our model is set up with Haar random unitaries, the approximation of \eqnref{eq:avgW1} by \eqnref{eq:avgW2} only requires local scrambling and should also hold for random Clifford circuits.  Since the Clifford group is a unitary 2-design \cite{DiVincenzo_2002}, this further implies that the dynamics of the purity for the Clifford circuit and Haar random circuit are identical within our formalism.}

Now the task is to evaluate the transfer matrix $\hat{T}_K$ for the quantum channel. Because \eqnref{eq:W=TW} is applicable to every Kraus operator $K_{x,t}$ in the quantum channel, $\hat{T}_{K}$ can be constructed from each single $\hat{T}_{K_{x,t}}$ recursively. Following \eqnref{eq:WK}, we find (see \appref{app:EFO} for derivation)
\begin{equation}\label{eq:TUTM}
\begin{split}
\hat{T}_{U_{ij}}&=\Big(1+\frac{d}{d^2+1}(X_i+X_j)\Big)\frac{1+Z_iZ_j}{2},\\
\hat{T}_{M_i}&=1-\frac{p}{d+1}+\frac{pd}{d+1}X_i,
\end{split}
\end{equation}
where $X_i$ and $Z_i$ denote the Pauli-$x$ and Pauli-$z$ operators acting on site $i$. Each step of the transfer matrix (see \figref{fig:circuit}) is then given by
\begin{equation}\label{eq:Tstep}
\hat{T}_\text{step}=\prod_{i}\hat{T}_{M_i}\prod_{\langle ij\rangle\in\text{even}}\hat{T}_{U_{ij}}\prod_{i}\hat{T}_{M_i}\prod_{\langle ij\rangle\in\text{odd}}\hat{T}_{U_{ij},}
\end{equation}
such that the full transfer matrix of $t$ steps (layers) of the quantum channel will be $\hat{T}_K=\hat{T}_\text{step}^t$. According to \eqnref{eq:W=TW}, the final entanglement feature state reads $\ket{W_{\rho}}=\hat{T}_\text{step}^t\ket{W_{\rho_0}}$, from which the 2nd R\'enyi entropy in the final state $\bar{\rho}$ can be retrieved based on \eqnref{eq:avgW2}, 
\begin{equation}\label{eq:S}
	S^{(2)}_{\bar{\rho}}[\sigma]=-\log W_{\bar{\rho}}[\sigma]\simeq -\log\frac{\braket{\sigma|W_{\rho}}}{\braket{\Uparrow|W_{\rho}}},
\end{equation} 
where the Ising configuration $[\sigma]$ labels the entanglement region. The denominator $\braket{\Uparrow|W_{\rho}}$ provides the appropriate normalization to ensure that the entanglement entropy vanishes for empty region, i.e.~$S_{\bar{\rho}}^{(2)}[\Uparrow]=0$. In the long-time limit ($t\to \infty$), the entanglement feature state $\ket{W_{\rho}}$ converges to the leading eigenvector of the one-step transfer matrix $\hat{T}_\text{step}$, denoted as $\ket{W_{\rho_\infty}}$.

\begin{figure}[htbp]
\begin{center}
\includegraphics[width=0.7\columnwidth]{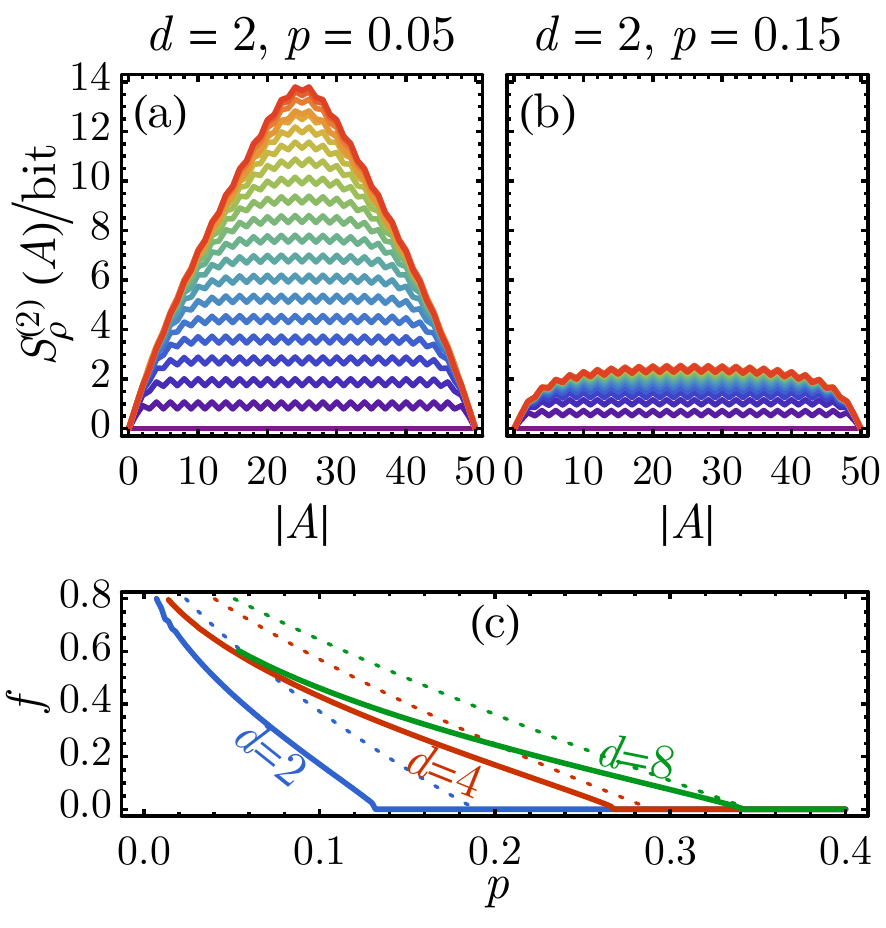}
\caption{Growth of entanglement entropy over a single region of size $|A|$ on a chain of 50 qubits in (a) the volume-law phase and (b) the area-law phase. The rainbow colors from purple to red correspond to the time step from 0 to 20. 
(c) The volume-law coefficient $f$ v.s. the measurement strength $p$ with different qudit dimensions $d$, where $f$ is extracted in the thermodynamic limit from the MPS representation of $\ket{W_{\rho_\infty}}$ with bond dimension 16. Dash lines are  upper bonds of $f$ by the quantum Hamming bound $d^{(1-f)N}\ge\left(\begin{smallmatrix}N\\pN\end{smallmatrix}\right) (d^2-1)^{\,pN}$.}
\label{fig:entropy}
\end{center}
\end{figure}

Driven by the measurement strength $p$, the leading eigenstate $\ket{W_{\rho_\infty}}$ can undergo a quantum phase transition that corresponds to the entanglement transition\cite{Vasseur2018Entanglement}. To see this, we need to calculate $\ket{W_{\rho_\infty}}$ for different $p$, which is still a challenging many-body problem. An important observation is that the entanglement feature state $\ket{W_{\rho}}$ itself is a low-entanglement state, even if its underlying physical quantum state $\rho$ can be highly entangled. Representing $\ket{W_{\rho}}$ as a matrix product state (MPS)\cite{Verstraete2008Matrix} enables us to tackle the problem using well-developed MPS-based numerical approaches\cite{Vidal2004Efficient,Verstraete2004Matrix,Zwolak2004Mixed-State,Zauner2018Variational} (see \appref{app:MPS} for algorithm details). We assume that the initial physical state $\rho_0$ is a random product state, whose entanglement feature state is $\ket{W_{\rho_0}}=\sum_{[\sigma]}\ket{\sigma}$, such that the entanglement entropy $S^{(2)}_{\rho_0}[\sigma]=0$ vanishes for all  entanglement regions. We numerically evolve $\ket{W_{\rho_0}}$ by $\hat{T}_\text{step}$ and present the growth and saturation of the entanglement entropy in \figref{fig:entropy}(a,b). We indeed observe the volume-law (area-law) behavior under small (large) measurement strength. Without the entanglement feature approach, it would be hard to directly simulate the volume-law state in \figref{fig:entropy}(a) with around 14 bits of half-system entanglement entropy. As the entanglement feature state converges to $\ket{W_{\rho_\infty}}$ in the long-time limit, we can extract the volume-law coefficient $f$, defined via $S^{(2)}_{\rho_\infty}(A)=(f\log d)|A|$. The result is shown in \figref{fig:entropy}(c), which clearly exhibits the measurement-driven entanglement transition for different qudit dimensions $d$, where different curves collapse to the same scaling form $f\log d \propto (p_c-p)^{\nu}$ with $\nu=1$ (see \appref{app:check}), implying the Ising universality class within the mean-field description. Nevertheless, the mean-field theory can not capture the universality correctly. Recent numerics indicate that the correct exponent $\nu$ should be $1.1\sim 1.3$\cite{Li2019METHQC,Gullans2019Dynamical,Choi2019QECEPTRUCWPM,Zabalo2019Critical}. 

\begin{figure}[htbp]
\includegraphics[width=0.84\columnwidth]{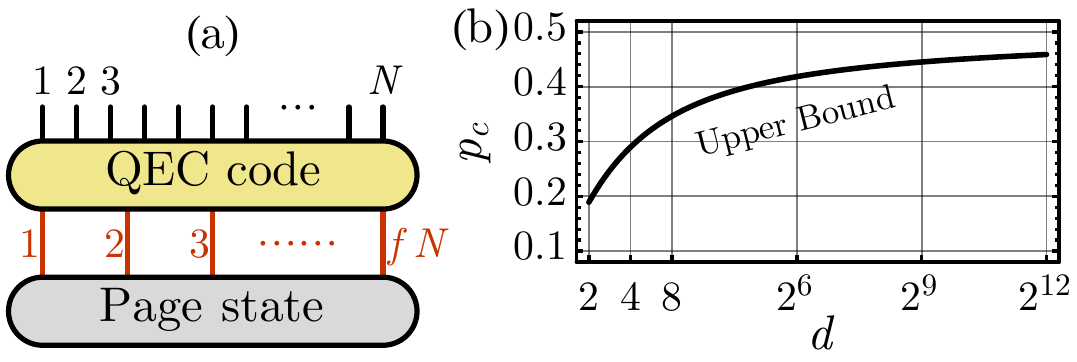}\\
\text{(c)}\begin{tabular}{|c|c|c|}
\hline
Dynamics & ${p_{c}}$ & Bound on ${p_{c}}$ \\ \hline
$\begin{array}{c} 
\text{Haar\,} (d=2) \\
 \text{Clifford\,} (d=2)
 \end{array}$ & $\begin{array}{c} 
0.168(5) \text{\,\,\cite{Zabalo2019Critical}} \\
0.16 \text{\,\,\cite{Li2019METHQC}}
 \end{array}$   & 0.1893 \\ \hline
 $d=\infty$ & 1/2 \text{\,\cite{Skinner2019MPTDE}} & 1/2\\\hline
\end{tabular}

\caption{(a) Assuming the final steady state can be modeled by a Page state on $f N$ qudits ($f\le 1$) encoded into a quantum error-correcting code on $N$ qudits, we find (b) a upper bound on the critical measurement strength $p_c$ for different qudit dimensions $d$ as in \eqnref{eq:pc_bound}, where $p_c\to1/2$ as $d\to\infty$. (c) Comparison with $p_c$ reported in literatures.}
\label{fig:QEC}
\end{figure}

\emph{Error Correcting Volume-Law States}--- The result in \figref{fig:entropy}(c) indicates that the volume-law phase is stable against finite strength of measurements. The volume-law scaling implies that the entropy associated with each qudit is $f\log d$ with $f \le 1$.  If a single-qudit measurement of strength $p$ reduced the qudit entropy by $p f\log d$, then after each layer of measurements, the entropy of a large region $A$ would be reduced in a volume-law manner $\Delta S^{(2)}(A)=-(p f\log d)|A|$, which is irremediable by the following layer of unitary gates, which only increases the entropy by an area-law amount $\Delta S^{(2)}(A)\propto|\partial A|\sim\scO(1)$. This would imply that the volume-law phase is unstable against  measurements, a paradox posted in \refcite{Chan2019UED}. It was pointed out in \refcite{Choi2019QECEPTRUCWPM} that the solution lies in the QEC \cite{Calderbank1996Good,Preskill1998Reliable} property in the sub-thermal volume-law state.
An example of such volume-law state on $N$ qudits can be obtained from encoding a Page state of $fN$ qudits by a layer of local QEC code as in \figref{fig:QEC}(a), which dilutes the Page state to a sub-thermal volume-law state with volume-law coefficient $f\le 1$. In each round of local measurements, $pN$ qudits will be measured typically, which effectively introduces errors up to weight $pN$. To prevent the measurement from disentangling the Page state and reducing the entanglement entropy extensively, the subsequent unitary layer should correct all errors (see \appref{app:QEC}). 
This requires the syndrome space dimension $d^{(1-f)N}$ to be at least as large as the number of error operators of weight $pN$,\footnote{This condition yields a so-called non-degenerate QEC code.  A similar bound on degenerate quantum codes -- were one to exist -- may improve the bound on the critical measurement probability presented in Eq. (\ref{eq:pc_bound}).} which yields the quantum Hamming bound\cite{Gottesman1996Class} $d^{(1-f)N}\ge\left(\begin{smallmatrix}N\\pN\end{smallmatrix}\right) (d^2-1)^{\,pN}$, see \figref{fig:entropy}(c).  The entanglement transition happens as $f \rightarrow 0$.  In the $N\rightarrow \infty$ limit, this gives a bound on the critical measurement rate $p_c$
\begin{equation}\label{eq:pc_bound}
p_{c}\log[(d^{2}-1)({p_{c}^{-1}-1})]\le \log[(1-p_{c})d],
\end{equation}
which is plotted in \figref{fig:QEC}(b).  For qubits ($d=2)$, this yields $p_{c} \le 0.1893$, the limit of infinite qudit dimension ($d\to\infty$) yields $p_{c} \le 1/2$, as summarized in \figref{fig:QEC}(c). The latter bound is saturated at the known transition point, corresponding to a bond percolation transition on the square lattice \cite{Skinner2019MPTDE,Bao2019TPTRUCWM,Jian2019MCRQC}.

\begin{figure}[htbp]
\includegraphics[width=0.7\columnwidth]{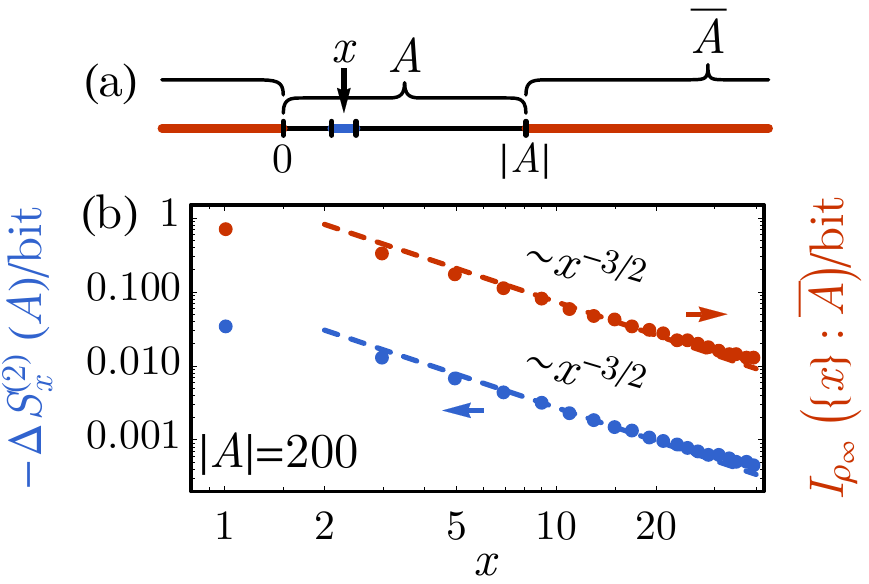}\\
\caption{(a) Entanglement region configuration for the mutual information $I_\rho(\{x\}:\bar{A})$. (b) Measurement-induced entropy drop $\Delta S_x^{(2)}(A)$ and the qubit-environment mutual information $I_{\rho_\infty}(\{x\}:\bar{A})$ for the final state of the random quantum channel (at $d=2, p=0.1$), based on the MPS of $\ket{W_{\rho_\infty}}$ with bond dimension 64. }
\label{fig:power}
\end{figure}

To quantify the QEC capacity in the sub-thermal volume-law state $\rho$ generated by the random quantum channel, we propose to study the mutual information $I_\rho(\{x\}:\bar{A})=S_{\bar{\rho}}^{(2)}(\{x\})+S_{\bar{\rho}}^{(2)}(\bar{A})-S_{\bar{\rho}}^{(2)}(\{x\}\cup\bar{A})$ between a qudit at $x$ (inside a region $A$) and the environment $\bar{A}$ (assuming $\bar{A}$ is larger than half of the system), see \figref{fig:power}(a). In terms of the entanglement feature state $\ket{W_\rho}$, we have (see \appref{app:entropy} for derivation)
\begin{equation}\label{eq:expI}
    e^{I_\rho(\{x\}:\bar{A})}=\frac{\braket{A|X_x|W_\rho}\braket{\Uparrow|W_\rho}}{\braket{A|W_\rho}\braket{\Uparrow|X_x|W_\rho}},
\end{equation}
where $\ket{A}=\prod_{i\in A}X_i\ket{\Uparrow}$ is the Ising basis state that specifies the region $A$. If $I_\rho(\{x\}:\bar{A})$ vanishes, measuring qudit $x$ in $A$ tells no information about $\bar{A}$, therefore the entanglement between $A$ and $\bar{A}$ is unaffected by the measurement, suggesting that the information about $\bar{A}$ has been scrambled in region $A$ to prevent local readout. It can be shown that the change of $S_\rho^{(2)}(A)$ after a measurement of strength $p$ at a qudit at $x$ distance away from the boundary of $A$ is directly related to $I_\rho(\{x\}:\bar{A})$ in the weak measurement limit $p\to0$ (see \appref{app:entropy}),
\begin{equation}\label{eq:DeltaS}
\begin{split}
    &\Delta S_x^{(2)}(A)\equiv-\log\frac{\braket{A|\hat{T}_{M_x}|W_\rho}}{\braket{\Uparrow|\hat{T}_{M_x}|W_\rho}}+\log\frac{\braket{A|W_\rho}}{\braket{\Uparrow|W_\rho}}\\
    &= 
    -\frac{pd}{d+1}W_{\bar{\rho}}(\{x\})\big(e^{I_{\rho}(\{x\}:\bar{A})}-1\big)+\scO(p^2)\,,
\end{split}
\end{equation}
where $W_{\bar{\rho}}(\{x\})$ is the single-qudit purity (at position $x$).
We found that the entropy drop depends on the measurement position $x$: a measurement deeper in the region $A$ will be less effective in reducing the entropy of $A$. Our MPS-based numerical calculation in \figref{fig:power}(b) confirms that $\Delta S_x^{(2)}(A)\sim -x^{-3/2}(|A|-x)^{-3/2}$ indeed follows the similar behavior as $I_\rho(\{x\}:\bar{A})$.\footnote{The proportionality constant is also consistent with the corresponding single qudit purity which is computed separately.} 
Both fall off with $x$ in a power-law manner with the exponent $3/2$. 
Given that the exponent $3/2$ is greater than $1$, the total entropy drop $\Delta S^{(2)}(A)=\sum_{x\in A}\Delta S_x^{(2)}(A)$ converges to a constant that does not scale with $|A|$, which can be balanced by the area-law entropy growth of the following unitary layer. Therefore the volume-law phase is stable.

To justify the exponent $3/2$, we approximate \footnote{Such approximation preserves the the long distance behavior as the entanglement transition is still described by the Ising universality class.}  the transfer matrices by $\hat{T}_{U_{ij}}\simeq e^{JZ_iZ_j}$ and $\hat{T}_{M_i}\simeq e^{hX_i}$. 
As $\hat{T}_{U_{ij}}$ ($\hat{T}_{M_i}$) can drive $\ket{W_\rho}$ towards the ferromagnetic (paramagnetic) state, the model still captures the volume-law (area-law) phase given $J>h$ ($J<h$). 
This simplification allows us to solve the leading eigenstate $\ket{W_{\rho_\infty}}$ of $\hat{T}_\text{step}$ analytically by mapping to the Majorana fermion basis $\chi_{2i-1}=\prod_{j<i}X_jZ_i$ and $\chi_{2i}=\prod_{j<i}X_jY_i$ by Jordan-Wigner transformation. In the fermion language, the entanglement feature $W_{\rho_\infty}(A)$ of a single region $A$ corresponds to a two-point strange correlator\cite{Shankar2011Equality,You2014Wave,Wierschem2016Detection} between free fermion states (see \appref{app:fermion})
\begin{equation}\label{eq:Winf_def}
    W_{\rho_\infty}(A)=\braket{A|W_{\rho_\infty}}=\braket{\Uparrow|\ii\chi_{0}\chi_{2|A|+1}|W_{\rho_\infty}},
\end{equation}
which was originally introduced to diagnose symmetry protected topological (SPT) orders. If $\ket{W_{\rho_\infty}}$ is in the topological (trivial) fermionic SPT phase (with respect to the reference state $\ket{\Uparrow})$\footnote{Strictly speaking, the reference state should be $\ket{\Uparrow}+\ket{\Downarrow}$ to preserve fermion parity, but the entanglement feature remains the same, given the $\ZZ_2$ symmetry for pure states.}, the strange correlator $W_{\rho_\infty}(A)$ will exhibit a long-range correlation (an exponential decay) with respect to $|A|$, matching the area-law (volume-law) entropy scaling. We calculated the strange correlator deep in the trivial phase with $h\ll J$ (see \appref{app:fermion}), and found
\begin{equation}\label{eq:Winf_asymptotic}
    W_{\rho_\infty}(A)\propto \frac{e^{-\kappa |A|}}{|A|^{3/2}},
\end{equation}
where $\kappa=\log(J/h)$. This unveils an important entanglement feature of the sub-thermal volume-law steady state $\rho_\infty$, namely the subleading logarithmic correction\cite{Li2019METHQC} of the single-region entanglement entropy $S_{\rho_\infty}^{(2)}(A)=\kappa |A|+\frac{3}{2}\log|A|$ with an universal coefficient $3/2$. The free fermion representation of $\ket{W_{\rho_\infty}}$ enables us to evaluate multi-region entanglement features as multi-point strange correlators, which can then be decomposed to two-point strange correlators using Wick's theorem. For example, the factor $\braket{A|X_x|W_{\rho_\infty}}=-\braket{\Uparrow|\chi_{0}\chi_{2x-1}\chi_{2x}\chi_{2|A|+1}|W_{\rho_\infty}}$ on the numerator of \eqnref{eq:expI} is a four-point correlator. Applying the asymptotic solution in \eqnref{eq:Winf_asymptotic}, we can confirm that the measurement-induced entropy drop $\Delta S_{x}^{(2)}(A)$ indeed decays with the measurement position as $x^{-3/2}$ with the universal exponent $3/2$, which is crucial to the stability of the volume-law phase. Finally, we show in Appendix \ref{app:qecc_argument} that the sub-thermal volume-law state generated by Clifford unitary gates and random measurements indeed exhibits a power-law dependence of the entanglement drop with the measurement position, due to the power-law dependence of the stabilizer length distribution in the steady-state \cite{Li2019METHQC}, though we are unable to derive the precise exponent appearing in the power-law decay (for the case of Clifford rather than Haar random unitaries). Our discussion reveals the QEC capacity of the sub-thermal volume-law state as a multi-region entanglement feature, which goes beyond the dichotomy of area-law v.s.~volume-law scaling of the single-region entanglement entropy, and demonstrates the advantage of  entanglement features in resolving finer structures of quantum many-body entanglement.

\emph{Order Parameter and Bulk Correlations}---
{ A natural question that arises is - how can we measure the $\mathbb{Z}_2$ Ising order parameter $\braket{Z}$ that appears within our mean-field description? In fact this is precisely the bulk order  parameter identified in  \cite{Zabalo2019Critical,Gullans2019Probes}, defined as the entanglement entropy of ancilla qudits, which are maximally entangled with the physical qudits during the circuit dynamics. 
The second R\'enyi entropy of a single ancilla is proportional to the bulk magnetization $\braket{Z}$. Clearly in the strong measurement phase, the ancilla is decoupled from the physical qudits and the order parameter vanishes. Also, the second R\'enyi mutual information between two (space-time) separated ancillas is proportional to the connected bulk two point correlation $\braket{Z_i Z_j}_c$\cite{Zabalo2019Critical,Gullans2019Probes}. Although the mean-field theory is not expected to correctly capture the critical fluctuations, nevertheless by way  of comparison we note that our Ising model mapping would imply, near the critical point, $\braket{Z}\propto (p_c-p)^{\beta}$ with $\beta=1/8$ and $\braket{Z_i Z_j}_c = |i-j|^{-\eta}$ with $\eta=1/4$, which, perhaps fortuitously, is close to the reported value in \cite{Zabalo2019Critical}.}

\emph{Acknowledgement}---
We acknowledge the helpful discussions with Matthew Fisher, John McGreevy, Tarun Grover, Xiao-Liang Qi, Ehud Altman, Xiao Chen, Yaodong Li, Yimu Bao, Shang Liu and Liujun Zou.  YZY acknowledges the previous collaborations with Yingfei Gu, Wei-Ting Kuo, Ahmed A. Akhtar, Daniel Arovas, Chao-Ming Jian, Romain Vasseur and Andreas W. W. Ludwig on relevant works.  SV is supported by the Harvard Society of Fellows. AV is supported by a Simons investigator Award and the DARPA DRINQs (award D18AC00033). This work was supported by the Simons Collaboration on Ultra-Quantum Matter, which is a grant from the Simons Foundation (651440, AV).

\bibliography{ref.bib}

\onecolumngrid
\newpage
\appendix
\setcounter{secnumdepth}{2}

\section{Consistency Checks}
\label{app:check}

In this section, we provide some consistency check for the two approximations we made in the main text. The first is the ``mean-field'' approximation as we explain below \eqnref{eq:avgW2}. The second is the ``Ising-model'' approximation $\hat{T}_{U_{ij}}\simeq e^{JZ_iZ_j},\,\hat{T}_{M_i}\simeq e^{hX_i}$ when justifying the exponent $3/2$.

\subsection{Check for the ``mean-field'' approximation}

As we have stated in the main text, most of the complication for analytical calculation comes from the renormalization of the wavefunction due to measurement, which is the denominator of \eqnref{eq:avgW1} and makes it hard for doing the average. If each measurement only reduces the norm of the wavefunction by a constant fraction, then the renormalization of the wavefunction can be absorbed into a redefinition of the measurement operator (which has been done appropriately in our definition) and one can effectively simplify \eqnref{eq:avgW1} to \eqnref{eq:avgW2}. In the following, we will show that this happens for the low measurement probability regime ($p\ll p_c$), which is exactly what we need for our later discussion for the volume law phase.

\begin{figure}[htbp]
    \centering
    \includegraphics[width=0.9\textwidth]{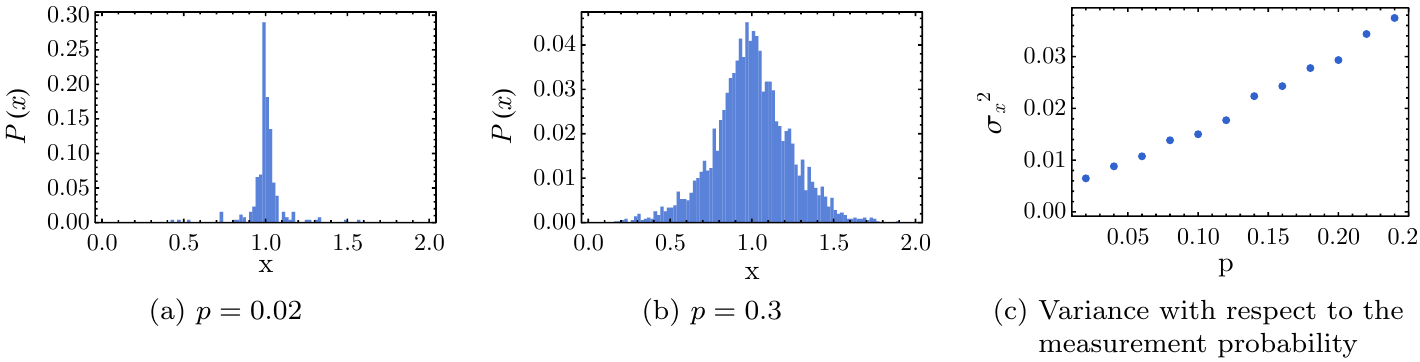}
    \caption{The statistics of the ratio of wavefunction norm square before and after measurements. For each plot, we simulate a 12 sites qubit chain, evolve it by 40 steps and repeat 80 times to do the average. Here each step consists of the two layers of unitary and measurements shown in \figref{fig:circuit}}
    \label{fig:check mean field}
\end{figure}

We use exact-diagonalization to implement the random quantum channel model shown in \figref{fig:circuit}, where each unitary is drawn from the Haar random unitary ensemble and the measurement is drawn from the ensemble $\{\II\}\cup\{\sqrt{d} P_V | V \in U(d) \}$ with the probability $P(\II) = 1-p$ and $P(\sqrt{d} P_V) = pdV$. We numerically study how the wavefunction norm square gets reduced by each measurement for different fixed measurement probability $p$. Namely, we collect the statistics of the following quantity
\begin{equation}
    x = \frac{\Tr \rho_{\text{after}}}{\Tr \rho_{\text{before}}}=\frac{\Tr M_i\rho_{\text{before}}M_i^\dagger}{\Tr \rho_{\text{before}}}\,,
\end{equation}
where $\rho_{\text{before}/\text{after}}$ represents the density matrix before and after each measurement respectively, such that $\rho_\text{after}=  M_i\rho_\text{before}M_i^\dagger$ with $M_i$ being a single-site measurement operator. The results are shown in \figref{fig:check mean field}. One can see from \figref{fig:check mean field}~(a) and (b) that the distribution of $x$ has a sharp peak around $x=1$ for small enough $p$ and gets broadened for larger $p$. \figref{fig:check mean field}~(c) directly shows that the variance of the distribution of $x$ monotonically decreases as we decrease $p$. That \figref{fig:check mean field}~(c) looks like a smooth curve without any discontinuity may be due to the small system size. Therefore, as long as we focus on the volume-law phase itself, the denominator of \eqnref{eq:avgW1} is almost like a constant and we can make the approximation. Nevertheless, the fact that $P(x)$ sharply peaks for small $p$ justifies that the fluctuation of $\Tr M_i\rho M_i^\dagger$ is strongly suppressed in the volume-law phase, in support of the ``mean-field'' approximation of replacing the average of ratio in \eqnref{eq:avgW1} by the ratio of averages in \eqnref{eq:avgW2}.

\subsection{Check for the ``Ising-model'' approximation}

\begin{figure}[htb]
    \centering
    \includegraphics[width=0.38\textwidth]{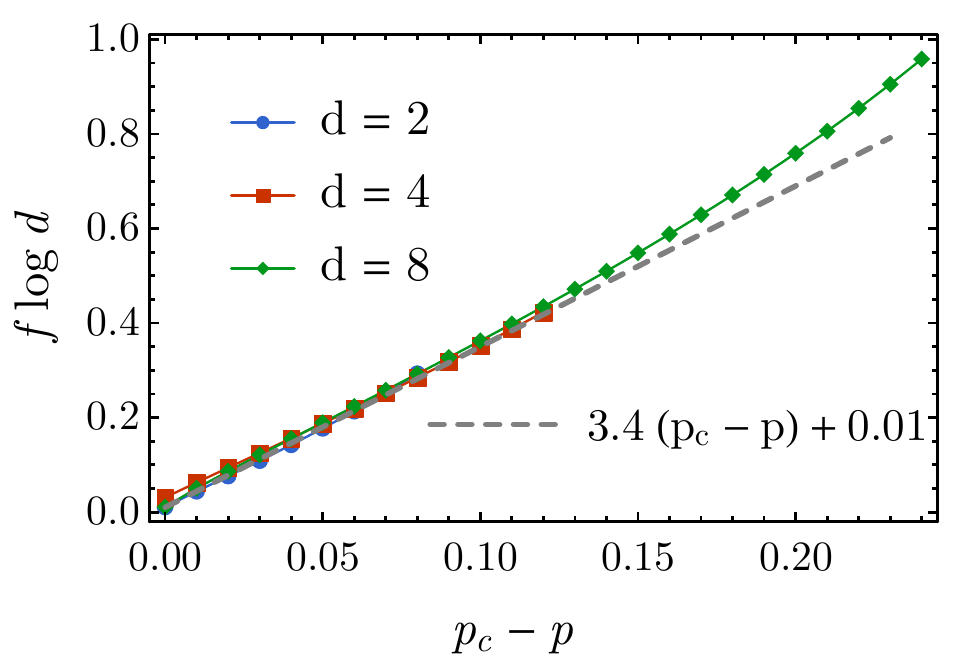}
    \caption{The volume-law coefficients with respect to the measurement probability for different qudit dimension. The horizontal axis is $p_c-p$ and the vertical axis is $f\log d$, for which $p_c$ and $\log d$ are chosen for different qudits respectively. All colored curves collapses for $p$ close to $p_c$. The gray curve is a linear function, which implies the critical exponent is $\nu=1$.}
    \label{fig:scaling collapse}
\end{figure}

As shown in \figref{fig:entropy}, the volume-law coefficient has a discontinuity at a certain critical value of $p_c$, which exhibits a phase transition. To verify that systems with different qudit dimensions $d$ share the same universality class of the transition, let us rescale the data and plot $f\log d$ as a function of $p_c-p$ for $p<p_c$ and results are in \figref{fig:scaling collapse}. Different curves collapsing with each other implies that they can be captured by the same scaling function $F((p_c-p)^\nu L)$. A further fitting yields that the critical exponent is $\nu=1$, which implies that the entanglement transition falls into the Ising universality class under the mean-field description. Although the actual universality class of the entanglement transition is beyond Ising, because the mean-field theory does not capture the critical fluctuation correctly, the result here is still meaningful in verifying that the entanglement dynamics can be approximated by a imaginary time Floquet problem of Ising model, see \eqnref{eqn:simplified Tstep}. Such approximation will not affect the long distance behavior and result in the same Ising universality class at the transition.

\section{Entanglement Feature Operators}
\label{app:EFO}
To construct the transfer matrix $\hat{T}_K$ of a Kraus operator $K$, we need to first calculate the corresponding entanglement feature operator $\hat{W}_K$. We direct the reader to \refcite{Kuo2019Markovian}, where the entanglement feature operator for the identity operator $\hat{W}_\id$ and the two-qudit Haar random unitary gate $\hat{W}_{U_{ij}}$ has been calculated. The result is
\begin{equation}
\begin{split}
\hat{W}_\id&=\prod_i d(d+X_i),\\
\hat{W}_{U_{ij}}&=d^2(d+X_i)(d+X_j)-\frac{d^2(d^2-1)}{2(d^2+1)}(1-Z_iZ_j)(d^2-X_iX_j).
\end{split}
\end{equation}
With these, we can already construct the transfer matrix for the unitary gate as
\begin{equation}\label{eq:TU}
\hat{T}_{U_{ij}}=\hat{W}_{U_{ij}}\hat{W}_{\id}^{-1}=\Big(1+\frac{d}{d^2+1}(X_i+X_j)\Big)\frac{1+Z_iZ_j}{2}.
\end{equation}

Here we derive the entanglement feature operator for the single-qudit measurement $M$, drawn from the ensemble $\scE_M=\{\id\}\cup\{\sqrt{d}P_V|V\in\U(d)\}$ (where $P_V=V\ket{0}\bra{0}V^\dagger$), equipped with the probability measure $P(\id)=1-p$ and $P(\sqrt{d}P_V)=p\dd V$. By definition
\begin{equation}
\begin{split}
W_{M}[\sigma,\tau]&=\mathop{\EE}_{M\in\scE_M}\Tr M^{\dagger\otimes 2}\scX_\sigma M^{\otimes 2}\scX_\tau\\
&=(1-p)\Tr \id^{\dagger\otimes 2}\scX_\sigma \id^{\otimes 2}\scX_\tau+p d^2\int_{\U(d)}\dd V \Tr P_V^{\dagger\otimes 2}\scX_\sigma P_V^{\otimes 2}\scX_\tau\\
&=(1-p)\Tr \scX_\sigma\scX_\tau+p d^2\int_{\U(d)}\dd V \Tr (\ket{0}\bra{0})^{\otimes 2}\scX_\sigma (\ket{0}\bra{0})^{\otimes 2}\scX_\tau\\
&=(1-p)d^{\frac{3+\sigma\tau}{2}}+pd^2.
\end{split}
\end{equation}
In terms of the operator form, we have
\begin{equation}
\hat{W}_M=\sum_{[\sigma,\tau]}\ket{\sigma}W_M[\sigma,\tau]\bra{\tau}=d^2+((1-p)d+ p d^2)X,
\end{equation}
from which the transfer matrix $\hat{T}_M$ can be constructed,
\begin{equation}\label{eq:TM}
\hat{T}_{M_i}=\hat{W}_{M_i}\hat{W}_{\id}^{-1}=1-\frac{p}{d+1}+\frac{d p}{d+1}X_i,
\end{equation}
where we have attached the site index $i$. Putting together \eqnref{eq:TU} and \eqnref{eq:TM}, we obtain the transfer matrices given in \eqnref{eq:TUTM}. The transfer matrix for each layer of the quantum channel can be further constructed out of these basic transfer matrices.

\section{Matrix Product State and Numerical Approaches}
\label{app:MPS}

\subsection{MPS Representation of Entanglement Feature State}
The entanglement feature state $\ket{W_\rho}$ was introduced to encode the entanglement feature of a many-body state $\rho$. But $\ket{W_\rho}$ itself is also a many-body state of Ising spins. We can further ask what is the entanglement property of $\ket{W_\rho}$? Is it an area-law state or a volume-law state? We do not have a full answer for this question in general, but the current understanding is that even the underlying physical state $\rho$ is volume-law entangled, its entanglement feature state $\ket{W_\rho}$ can still be area-law entangled. This can be shown by an explicit construction of the matrix product state (MPS) representation for the entanglement feature state of the Page state (which is an extreme limit of the volume-law state with maximal thermalization). Let us consider the following translational invariant MPS ansatz for the entanglement feature
\begin{equation}\label{eq:WMPS1}
W_\rho[\sigma]=\Tr(\cdots A^{\sigma_{i-1}}A^{\sigma_{i}}A^{\sigma_{i+1}}\cdots),
\end{equation}
where $A^{\sigma}$ is a matrix specified by the Ising spin $\sigma=\pm1$. We claim that the following setting of $A^\sigma$ gives an exact MPS representation (up to a normalization constant) for the entanglement feature of the Page state
\begin{equation}\label{eq:APage}
A^\sigma=\mat{d^{\sigma/2}&0\\0&d^{-\sigma/2}}.
\end{equation}
Plugging \eqnref{eq:APage} to \eqnref{eq:WMPS1}, we can show
\begin{equation}
\begin{split}
W_\rho[\sigma]&=\Tr\prod_i\mat{d^{\sigma_i/2}&0\\0&d^{-\sigma_i/2}}=d^{\frac{1}{2}\sum_{i}\sigma_i}+d^{-\frac{1}{2}\sum_{i}\sigma_i},\\
S_{\bar{\rho}}^{(2)}[\sigma]&=-\log\frac{W_\rho[\sigma]}{W_\rho[\Uparrow]}=-\log\frac{d^{\frac{1}{2}\sum_{i}\sigma_i}+d^{-\frac{1}{2}\sum_{i}\sigma_i}}{d^{\frac{N}{2}}+d^{-\frac{N}{2}}}.
\end{split}
\end{equation}
This precisely matches the entanglement feature of the Page state for $N$ qudits (each of the dimension $d$). It produces the volume-law entanglement entropy scaling with maximal volume-law coefficient $f=1$. So the Page state entanglement feature admits an MPS representation of bond dimension 2. On the other hand, the product state entanglement feature $W_\rho[\sigma]=1$ can obviously be produced by an even simpler ansatz $A^\sigma=\mat{1}$, which is of the bond dimension 1. We can see, both the unentangled and maximally-entangled limit of the entanglement feature can be captured by MPS with low bond dimension. It is conceivable that the MPS ansatz may provide pretty good description for intermediate states across the entanglement transition as well. It is also expected that the MPS description will fall short at the transition: as $\ket{W_\rho}$ becomes critical, the required MPS bond dimension scales with the system size logarithmically.

We use two MPS-based numerical approaches in this work: the time-evolving block decimation (TEBD) algorithm\cite{Vidal2004Efficient,Verstraete2004Matrix,Zwolak2004Mixed-State} and the variational uniform matrix product state (VUMPS) algorithm\cite{Zauner2018Variational}. We use the TEBD algorithm to evolve the entanglement feature state $\ket{W_\rho}$ in time following entanglement dynamics specified by the random quantum channel model. We use the VUMPS to find the final entanglement feature state $\ket{W_{\rho_\infty}}$ in the long-time limit (as the leading eigenstate of the transfer matrix).

\subsection{TEBD Approach}
We first introduce the TEBD approach. We study the entanglement dynamics under the random quantum channel model. The evolution of the entanglement feature state $\ket{W_\rho}\to\hat{T}_K\ket{W_\rho}$ is governed by the transfer matrix $\hat{T}_K$ of the quantum channel $K$, 
\begin{equation}\label{eq:TKbrickwall}
\hat{T}_K=\dia{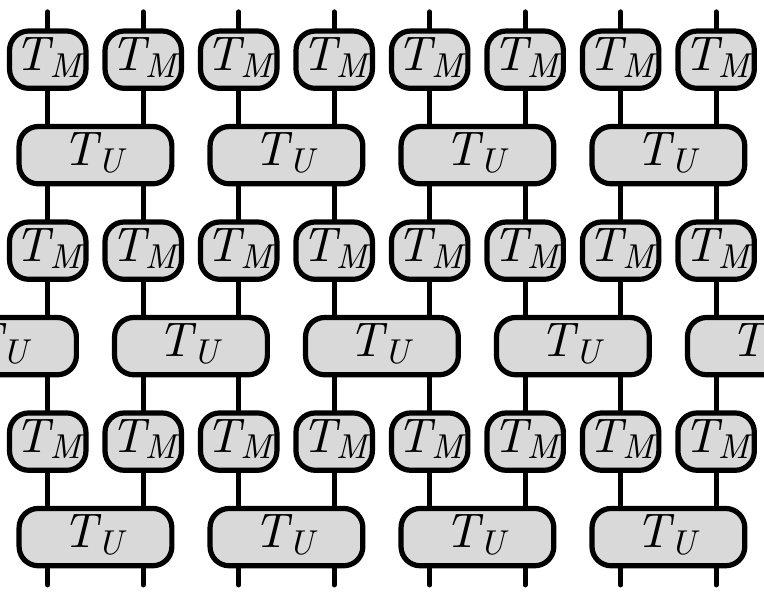}{100}{-50}
\end{equation}
which consists of the transfer matrix $\hat{T}_{U_{ij}}$ for the two-qudit unitary gate $U_{ij}$ and the transfer matrix $\hat{T}_{M_i}$ for the single-qudit weak measurement $M_i$. They are arranged in the brick-wall pattern as shown in \eqnref{eq:TKbrickwall}. Their expressions are given in \eqnref{eq:TU} and \eqnref{eq:TM} respectively. We start with the entanglement feature state of product states $\ket{W_\rho}=\sum_{[\sigma]}\ket{\sigma}$, which is translation invariant. Because the transfer matrix has a 2-site translation symmetry, we expect that the resulting entanglement feature state will also respect the 2-site translation symmetry, and can be described by an MPS ansatz with 2-site unit-cells,
\begin{equation}
W_\rho[\sigma]=\Tr\Big(\prod_j A_1^{\sigma_{2j-1}}A_2^{\sigma_{2j}}\Big)=\Tr(\cdots\dia{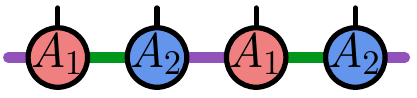}{18}{-5}\cdots).
\end{equation}
The MPS tensors are initialized to
\begin{equation}\label{eq:A1A2init}
A_1^\sigma=A_2^\sigma=\mat{1},
\end{equation}
which parameterizes the entanglement feature of product states. We then apply the TEBD algorithm, as described in Algorithm~\ref{alg:TEBD}, to evolve the MPS representation of $\ket{W_\rho}$ in time, where transfer matrices $\hat{T}_U$ and $\hat{T}_M$ are applied to $\ket{W_\rho}$ step-by-step following \eqnref{eq:TKbrickwall}.

\begin{breakablealgorithm}
\caption{Applying TEBD to evolve the MPS of $\ket{W_\rho}$}\label{alg:TEBD}
\begin{flushleft}
        \textbf{input:} $\hat{T}_U=\dia{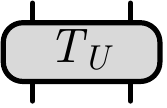}{18}{-7}$, $\hat{T}_M=\dia{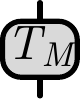}{18}{-7}$ - transfer matrices of two-qudit gate $\hat{T}_U$ and single-qudit measurement $\hat{T}_M$.\\
        \textbf{output:} $\ket{W_\rho}=\Tr(\cdots\dia{MPS1}{17}{-4}\cdots)$ - MPS representation of the entanglement feature state after $T$ steps of evolution (following the brick-wall circuit).
\end{flushleft}
\begin{algorithmic}[1]
	\Procedure{TEBD}{$T$}
		\State $\dia{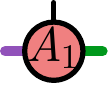}{16}{-4}\gets\mat{\mat{1},\mat{1}},\dia{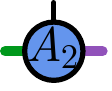}{16}{-4}\gets\mat{\mat{1},\mat{1}},\dia{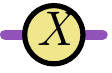}{13}{-4}\gets\mat{1}$ \Comment initialization (start with $\ket{W_\rho}$ of product states)
		\For{$t=1:2T$} \Comment evolves for $T$ steps
			\State $(\dia{A1}{16}{-4},\dia{A2}{16}{-4},\dia{X1}{13}{-4})\gets\text{TEBD.iterate}(\dia{A1}{16}{-4},\dia{A2}{16}{-4},\dia{X1}{13}{-4})$
		\EndFor
	\State\Return$(\dia{A1}{16}{-4},\dia{A2}{16}{-4})$
	\EndProcedure
	\vspace{8pt}
	\Function{TEBD.iterate}{$A_1, A_2, X$}
		\State $(\dia{A1}{16}{-4},\dia{A2}{16}{-4},\dia{X1}{13}{-4})\gets (A_1, A_2, X)$ \Comment import MPS tensors $A_1,A_2$ and symmetry operator $X$
		\State $\dia{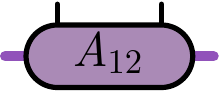}{16}{-4} \gets \dia{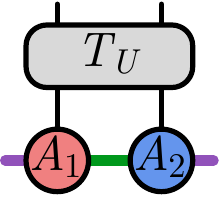}{35}{-4}$ \Comment apply transfer matrix $\hat{T}_U$
		\State $\dia{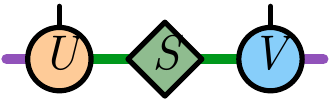}{18}{-6}\gets \text{SVD}(\dia{A12}{16}{-4},\text{up to }D_{cut})$ \Comment perform SVD up to cutoff dimension $D_{cut}$
		\State $\dia{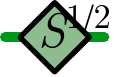}{14}{-5}\gets\text{sqrt}(\dia{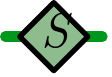}{14}{-5}/\text{max}(\dia{S}{14}{-5}))$ \Comment normalize singular values and take square root
		\State $(\dia{A1}{16}{-4},\dia{A2}{16}{-4})\gets (\dia{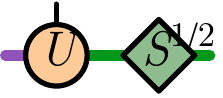}{17}{-5},\dia{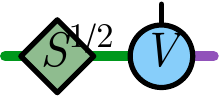}{17}{-5})$ \Comment construct new MPS tensors
		\State $(\dia{A1}{16}{-4},\dia{A2}{16}{-4})\gets (\dia{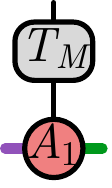}{35}{-4},\dia{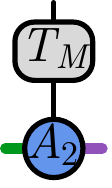}{35}{-4})$ \Comment apply transfer matrix $\hat{T}_M$
		\State $\dia{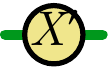}{13}{-5}\gets\text{round}(\dia{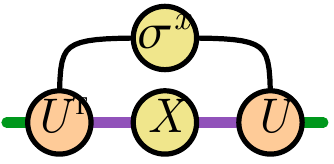}{28}{-5})$ \Comment construct new $\ZZ_2$ symmetry operator
		\State $(\dia{A1}{16}{-4},\dia{A2}{16}{-4})\gets \frac{1}{2}(\dia{A1}{16}{-4}+\dia{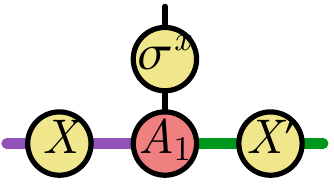}{32}{-5},\dia{A2}{16}{-4}+\dia{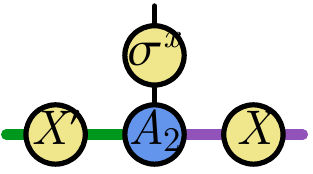}{32}{-5})$ \Comment impose $\ZZ_2$ symmetry (by symmetrization)
		\State\Return $(\dia{A2}{16}{-4},\dia{A1}{16}{-4},\dia{X2}{13}{-4})$ \Comment return with $A_1,A_2$ switched
	\EndFunction
\end{algorithmic}
\end{breakablealgorithm}

One important point is to preserve the $\ZZ_2$ symmetry under the evolution. The entanglement feature for pure states is $\ZZ_2$ symmetric, i.e.~$W_\rho[\sigma]=W_\rho[-\sigma]$. The symmetry acts on the MPS tensors $A_1$ and $A_2$ as
\begin{equation}
\dia{A1}{16}{-4}\to\dia{A1XXX}{32}{-5},\quad \dia{A2}{16}{-4}\to\dia{A2XXX}{32}{-5},
\end{equation}
where $X$ and $X'$ are representations of the $\ZZ_2$ symmetry operator in MPS auxiliary spaces. They must be updated in each iteration with the MPS tensor. Initially, we start with
\begin{equation}
X=\mat{1},
\end{equation}
which is consistent with the initial setup of $A_1,A_2$ in \eqnref{eq:A1A2init}. As new auxiliary degrees of freedom emerge under the singular value decomposition, the $\ZZ_2$ symmetry action should be calculated. The idea is to transform the $\ZZ_2$ symmetry action on the old degrees of freedom to the new degrees of freedom by the isometry constructed in SVD. We can show that the following two constructions are equivalent (assuming that the singular values have no accidental degeneracy)
\begin{equation}
\dia{X2}{13}{-5}=\dia{UXU}{28}{-5}=\dia{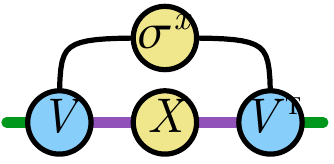}{28}{-5}.
\end{equation}
This is the step taken in line 15 of Algorithm~\ref{alg:TEBD}. The additional round off function is applied to eliminate numerical error accumulated in the calculation, so as to obtain a precise $\ZZ_2$ symmetry operator $X'$ which squares to identity $X'^2=\id$ precisely. The symmetry is implemented at each iteration by symmetrizing the MPS tensors $A_1,A_2$ as shown in line 16 of Algorithm~\ref{alg:TEBD}.

As we obtain the MPS tensors $A_1,A_2$ after $2T$ steps of the TEBD iteration (two TEBD iteration correspond to one step of time-evolution in the quantum channel model), we can calculate the entanglement entropy from the entanglement feature $S_{\bar{\rho}}^{(2)}[\sigma]=-\log(W_\rho[\sigma]/W_\rho[\Uparrow])$. In particular, if we consider a single entanglement region $A$ of size $|A|$ in a system of $N$ qudits, the entanglement entropy is given by
\begin{equation}
S_{\bar{\rho}}^{(2)}(A)=\left\{
\begin{array}{ll}
-\log\frac{\Tr(A_1^\downarrow A_2^\downarrow)^{|A|/2}(A_1^\uparrow A_2^\uparrow)^{(N-|A|)/2}}{\Tr(A_1^\uparrow A_2^\uparrow)^{N/2}} & |A|\in\text{even},\\
-\log\frac{\Tr(A_1^\downarrow A_2^\downarrow)^{(|A|-1)/2}A_1^\downarrow A_2^\uparrow(A_1^\uparrow A_2^\uparrow)^{(N-|A|-1)/2}}{\Tr(A_1^\uparrow A_2^\uparrow)^{N/2}} & |A|\in\text{odd}.\\
\end{array}\right.
\end{equation}
We follow this approach to calculated the entropy growth in \figref{fig:entropy}(a,b). The calculation is done with the MPS bond dimension cutoff at 64.

\subsection{VUMPS Approach}
In principle, if we follow the TEBD iteration for infinite steps, the MPS should converge to the leading eigenstate $\ket{W_{\rho_\infty}}$ of the transfer matrix. However, the TEBD algorithm is not stable under long-time evolution, as the error rate can not go down due to the SVD truncation at each iteration, hence TEBD is not good for targeting the final state $\ket{W_{\rho_\infty}}$. The VUMPS algorithm was proposed to avoid SVD truncation by variational optimization. To proceed, we first rewrite the transfer matrix into a matrix product operator (MPO) form. We notice that the transfer matrix $\hat{T}_K$ in \eqnref{eq:TKbrickwall} can be deformed to the following form
\begin{equation}
\hat{T}_K=\dia{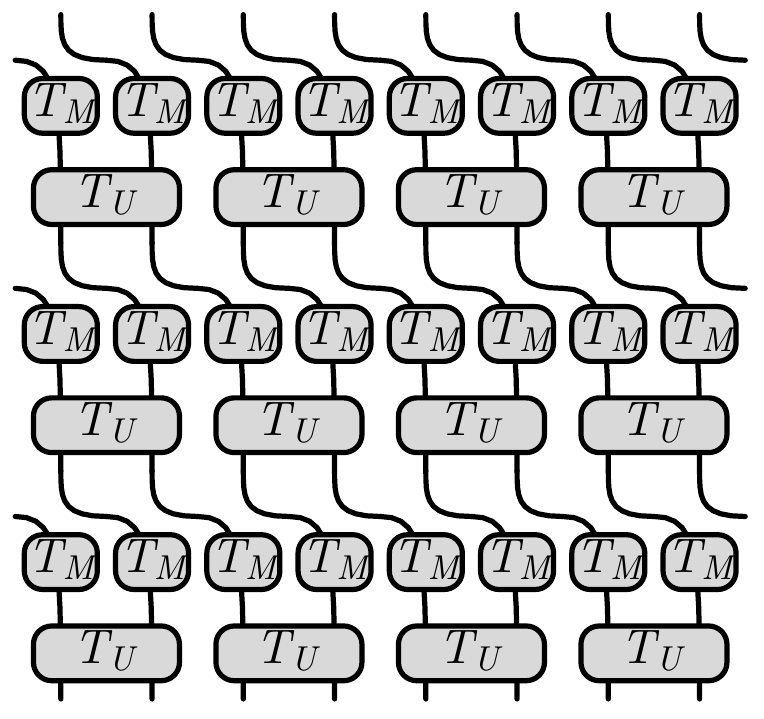}{120}{-60},
\end{equation}
such that the network acquires a one-layer translation symmetry along the time direction. Thus we introduce the single-layer transfer matrix $\hat{T}_\text{layer}$,
\begin{equation}
\hat{T}_\text{layer}=\cdots\dia{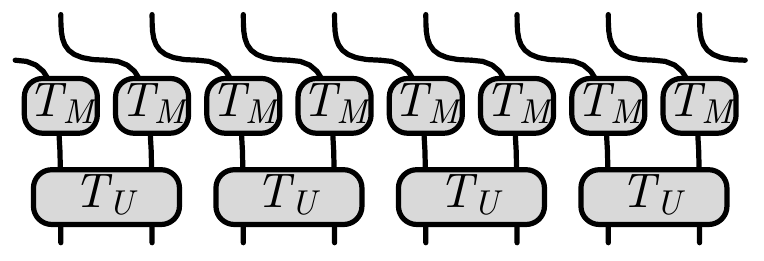}{42}{-30}\cdots,
\end{equation}
such that $\hat{T}_K=\hat{T}_\text{layer}^{2T}$ for $T$ steps of evolution. We further notice that each $\hat{T}_{U_{ij}}$ operator comes with a projection operator $(1+Z_iZ_j)/2$, such that only the $Z_{2j-1}Z_{2j}=+1$ states can survive the projection across neighboring layers. Thus we can restrict ourselves to the subspace of $\forall j: Z_{2j-1}Z_{2j}=+1$ and simplify the transfer matrix $\hat{T}_\text{layer}$ to
\begin{equation}
\begin{split}
\hat{T}_\text{layer}&=\cdots\dia{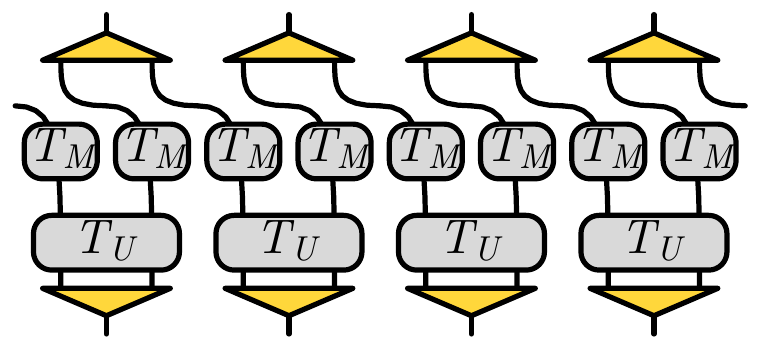}{56}{-37}\cdots\\
&=\cdots\dia{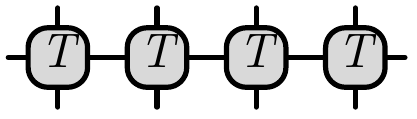}{20}{-8}\cdots,
\end{split}
\end{equation}
where each yellow triangle denotes a projection operator that projects to the $Z_{2j-1}Z_{2j}=+1$ subspace. In this way, the layer transfer matrix $\hat{T}_\text{layer}$ can be written as an MPO, with the MPO tensor given by
\begin{equation}
\dia{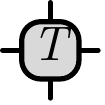}{18}{-7}=\dia{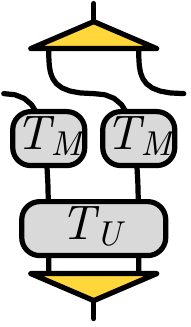}{56}{-37}.
\end{equation}
Arranging the legs following the order of up, down, left and right, the four-leg MPO tensor $T$ can be represented in the following tensor form
\begin{equation}
T=\mat{\mat{a&0\\c&0}&\mat{b&0\\c&0}\\\\ \mat{0&c\\0&b}&\mat{0&c\\0&a}},
\end{equation}
with tensor elements specified by
\begin{equation}
\begin{split}
a=\frac{(d+1-p)(d^2+(d-1)p+1)}{(d+1)(d^2+1)},\\
b=\frac{d^2 p((d-1)p+2)}{(d+1)(d^2+1)},\\
c=\frac{d((d-1)p+1)}{d^2+1},\\
\end{split}
\end{equation}
where $d$ is the qudit dimension and $p$ is the measurement strength. They are the only two tuning parameters of the random quantum channel model. Having specified the MPO tensor $T$, we can find the MPS representation of the leading eigenstate $\ket{W_{\rho_\infty}}$ of the layer transfer matrix $\hat{T}_\text{layer}$ using the VUMPS algorithm as described in Algorithm~\ref{alg:VUMPS}.

\begin{breakablealgorithm}
\caption{Applying VUMPS to find the MPS of the leading eigenstate $\ket{W_{\rho_\infty}}$}\label{alg:VUMPS}
\begin{flushleft}
        \textbf{input:} $\hat{T}_\text{layer}=\cdots\dia{MPO}{20}{-8}\cdots$ - MPO representation of the layer transfer matrix.\\
        \textbf{output:} $\ket{W_{\rho_\infty}}=\cdots\dia{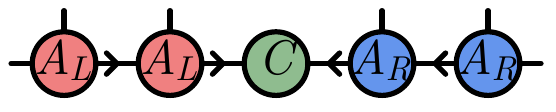}{19}{-6}\cdots$ - canonicalized MPS representation of the leading eigenstate of the layer transfer matrix.
\end{flushleft}
\begin{algorithmic}[1]
	\Procedure{VUMPS}{}
		\State $\dia{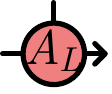}{16}{-4}\gets\dia{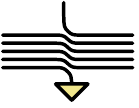}{18}{-7},\dia{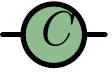}{13}{-4}\gets\dia{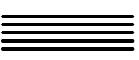}{10}{-2},\dia{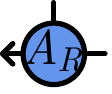}{16}{-4}\gets\dia{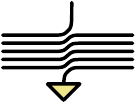}{18}{-7}$ \Comment initialize MPS tensors
		\While{$(A_L,C,A_R)$ not converge} \Comment iterate to improve MPS tensors
			\State $(\dia{AL}{16}{-4},\dia{C}{13}{-4},\dia{AR}{16}{-4})\gets\text{VUMPS.iterate}(\dia{AL}{16}{-4},\dia{C}{13}{-4},\dia{AR}{16}{-4})$
		\EndWhile
	\State\Return$(\dia{AL}{16}{-4},\dia{C}{13}{-4},\dia{AR}{16}{-4})$
	\EndProcedure
	\vspace{8pt}
	\Function{VUMPS.iterate}{$A_L, C, A_R$}
		\State $(\dia{AL}{16}{-4},\dia{C}{13}{-4},\dia{AR}{16}{-4})\gets (A_L,C,A_R)$ \Comment import MPS tensors
		\While{$(T_L,T_R)$ not converge} 
			\State $\dia{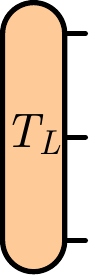}{48}{-21}\gets\text{normalize}\Big(\dia{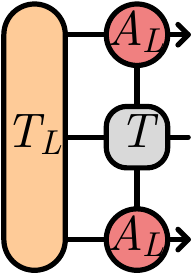}{48}{-21}\Big)$ \Comment power iteration to find the leading left-environment tensor $T_L$
			\State $\dia{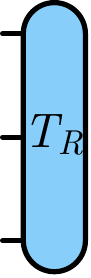}{48}{-21}\gets\text{normalize}\Big(\dia{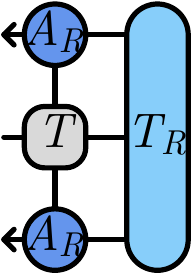}{48}{-21}\Big)$ \Comment power iteration to find the leading right-environment tensors $T_R$
		\EndWhile
		\While{$(C,B)$ not converge} 
			\State $\dia{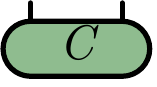}{16}{-4}\gets\text{normalize}\Big(\dia{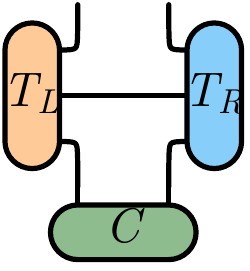}{48}{-21}\Big)$ \Comment power iteration to find the leading MPS central tensor $C$
			\State $\dia{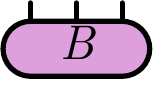}{16}{-4}\gets\text{normalize}\Big(\dia{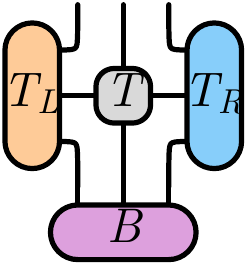}{48}{-21}\Big)$ \Comment power iteration to find the leading MPS block tensor $B$
		\EndWhile
		\State $\dia{AL}{16}{-4}\gets\text{minimize}(\norm{\dia{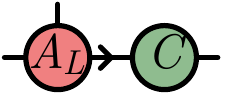}{16}{-4}-\dia{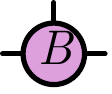}{16}{-4}},\text{ subject to }\dia{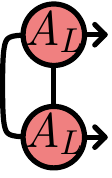}{30}{-13}=\dia{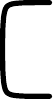}{18}{-7})$ \Comment optimize the left-isometry tensor $A_L$
		\State $\dia{AR}{16}{-4}\gets\text{minimize}(\norm{\dia{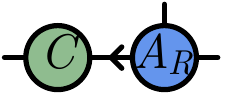}{16}{-4}-\dia{B}{16}{-4}},\text{ subject to }\dia{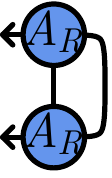}{30}{-13}=\dia{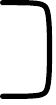}{18}{-7})$ \Comment optimize the right-isometry tensor $A_R$
		\State $(\dia{AL}{16}{-4},\dia{C}{13}{-4},\dia{AR}{16}{-4})\gets \frac{1}{2}(\dia{AL}{16}{-4}+\dia{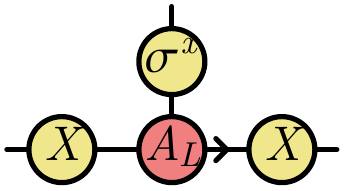}{32}{-5},\dia{C}{13}{-4}+\dia{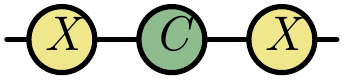}{14}{-5},\dia{AR}{16}{-4}+\dia{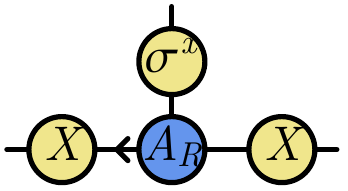}{32}{-5})$ \Comment impose $\ZZ_2$ symmetry (by symmetrization)
		\State\Return$(\dia{AL}{16}{-4},\dia{C}{13}{-4},\dia{AR}{16}{-4})$
	\EndFunction
\end{algorithmic}
\end{breakablealgorithm}

The VUMPS works with a canonicalized MPS, meaning that the MPS consists of left-isometry tensors $A_L$, right-isometry tensors $A_R$ and a central tensor $C$, as follows
\begin{equation}
\ket{W_{\rho_\infty}}=\cdots\dia{MPS2}{19}{-6}\cdots,
\end{equation}
where little arrows mark the direction of isometry map (mapping from large space into smaller space). The isometry tensors are initialized to
\begin{equation}
\dia{AL}{16}{-4}=\dia{ALinit}{18}{-7},\quad \dia{AR}{16}{-4}=\dia{ARinit}{18}{-7},
\end{equation}
where each thin line denotes a 2-dimensional space (i.e.~a qubit). The auxiliary space contains $n$ qubits and is of the dimension $2^n$, where $n$ is a hyper-parameter that can be adjusted. Larger $n$ (larger bond dimension) will generally result in better MPS representation. The isometry tensors initially collect the physical legs of $n$ MPO tensors $T$ away from the center. The little yellow triangle is taken to be a $\ZZ_2$ symmetric qubit state $(\ket{0}+\ket{1})/\sqrt{2}$, which is introduced to ``ground'' the physical legs of MPO tensors more than $n$ steps away from the center. The central tensor $C$ is simply taken to be an identity operator in the $n$-qubit auxiliary space. The initial ansatz is such chosen to preserve the $\ZZ_2$ symmetry from the beginning. The symmetry acts on the tensors as
\begin{equation}
\dia{AL}{16}{-4}\to\dia{ALXXX}{32}{-5},\quad \dia{C}{13}{-4}\to\dia{XCX}{14}{-5},\quad \dia{AR}{16}{-4}\to\dia{ARXXX}{32}{-5},
\end{equation}
where $X=\prod_{i=1}^n\sigma_i^x$ is the representation of the $\ZZ_2$ symmetry operator in the auxiliary space. The operator $X$ is fixed under VUMPS iteration, because VUMPS is a variational approach which does not reshuffle existing basis or generate new basis. We impose the $\ZZ_2$ symmetry by explicit symmetrization in line 20 of Algorithm~\ref{alg:VUMPS}. A key step in the algorithm is to efficiently reconstruct $A_L, A_R$ by solving the optimization problem in line 18, 19 of Algorithm~\ref{alg:VUMPS}. We direct the reader to \refcite{Zauner2018Variational} for details about how the solution can be approximately constructed in a robust manner.

\subsection{Extracting Volume-Law Coefficient}
As the VUMPS iteration converges, we obtain the tensors $A_L$, $C$ and $A_R$ which are needed to construct the canonicalized MPS state $\ket{W_{\rho_\infty}}$. We can then study all entanglement features of the final state produced by the random quantum channel. In particular, we can extract the volume-law coefficient $f$ which is defined via the scaling of entanglement entropy $S_{\bar{\rho}_\infty}^{(2)}(A)=(f\log d) |A|+\cdots$ in the $|A|\to\infty$ limit. We first solve the eigen problem of $A_L^\uparrow$ and $A_R^\uparrow$ (note that the isometry is assumed to go from the column space to the row space for $A_L^\uparrow$ and $A_R^\uparrow$),
\begin{equation}
\begin{split}
A_L^\uparrow\ket{\lambda_{Lm}}&=\lambda_{Lm}\ket{\lambda_{Lm}},\\
A_R^\uparrow\ket{\lambda_{Rm}}&=\lambda_{Rm}\ket{\lambda_{Rm}},
\end{split}
\end{equation}
where $m=0,1,2,\cdots$ labels the eigenvalues in a descending order $\lambda_{L0}\geq\lambda_{L1}\geq\lambda_{L2}\geq\cdots$. In fact, only the first two eigenvalues will be needed. Due to the $\ZZ_2$ symmetry, the eigenstates of $A_L^\downarrow$ ($A_R^\downarrow$) are related to that of $A_L^\uparrow$ ($A_R^\uparrow$) as $X\ket{\lambda_{Lm}}$ ($X\ket{\lambda_{Rm}}$) by applying the symmetry operator $X$, and the corresponding eigenvalues must be the same. There is also a reflection symmetry about the center, which relates the eigenvalues between $A_L^\uparrow$ and $A_R^\uparrow$ such that $\lambda_{Lm}=\lambda_{Rm}=\lambda_{m}$. Numerically there is often a slight difference between $\lambda_{Lm}$ and $\lambda_{Rm}$ due to the numerical error, so we define $\lambda_m=\sqrt{\lambda_{Lm}\lambda_{Lm}}$ as their geometric mean in practice. Given the setup, we can evaluate the entanglement feature for a region $A$ of size $|A|$ in a system of $N$ qudits,
\begin{equation}\label{eq:WinfMPS1}
W_{\rho_\infty}(A)=\bra{+}(A_L^{\downarrow\intercal})^{|A|/2}C(A_R^\uparrow)^{(N-|A|)/2}\ket{+},
\end{equation}
where $\ket{+}$ specifies the boundary condition for the MPS. The choice of $\ket{+}$ will not be important in the thermodynamic limit (as $|A|,N\to\infty$), because only the leading eigenstate dominates in the end. We only require $\ket{+}$ to be a $\ZZ_2$ symmetric state (i.e.~$X\ket{+}=\ket{+}$). For example, $\ket{+}=(1+X)\ket{0}$ is a possible choice. Using the $\ZZ_2$ symmetry property $A_L^\downarrow=XA_L^\uparrow X$ and $X^2=\id$, \eqnref{eq:WinfMPS1} can be written as
\begin{equation}\label{eq:WinfMPS2}
\begin{split}
W_{\rho_\infty}(A)&=\bra{+}(XA_L^{\uparrow\intercal}X)^{|A|/2}C(A_R^\uparrow)^{(N-|A|)/2}\ket{+}\\
&=\bra{+}X(A_L^{\uparrow\intercal})^{|A|/2}XC(A_R^\uparrow)^{(N-|A|)/2}\ket{+}\\
&=\bra{+}(A_L^{\uparrow\intercal})^{|A|/2}XC(A_R^\uparrow)^{(N-|A|)/2}\ket{+}.
\end{split}
\end{equation}
Here we have assumed that both $|A|$ and $N$ are even in sites, which means that they are integer in unit-cells. In this way, the entanglement cut will always pass between unit-cells, which simplifies our calculation. For the purpose of calculating the volume-law coefficient, such choice of entanglement cut does not affect the result. Suppose the state $\ket{+}$ admits the following decomposition $\ket{+}=\sum_m c_{Lm}\ket{\lambda_{Lm}}=\sum_m c_{Rm}\ket{\lambda_{Rm}}$ on the eigenstates with some (unimportant) coefficients $c_{Lm}$ and $c_{Rm}$, then \eqnref{eq:WinfMPS2} becomes
\begin{equation}\label{eq:WinfMPS3}
W_{\rho_\infty}(A)=\sum_{m,m'}c_{Lm}c_{Rm'}\lambda_{m}^{|A|/2}\lambda_{m'}^{(N-|A|)/2}\braket{\lambda_{Lm}|XC|\lambda_{Rm'}}.
\end{equation}
The entanglement entropy is given by
\begin{equation}
S_{\bar{\rho}_\infty}^{(2)}(A)=-\log\frac{W_{\rho_\infty}(A)}{W_{\rho_\infty}(\emptyset)}.
\end{equation}
We are interested in its slope with respect to $|A|$, thus we take the derivative
\begin{equation}\label{eq:dS1}
\begin{split}
\partial_{|A|}S_{\bar{\rho}_\infty}^{(2)}(A)&=-\frac{\partial_{|A|}W_{\rho_\infty}(A)}{W_{\rho_\infty}(A)}\\
&=-\frac{\sum_{m,m'}c_{Lm}c_{Rm'}\lambda_{m}^{|A|/2}\lambda_{m'}^{(N-|A|)/2}(\log\lambda_{m}-\log\lambda_{m'})\braket{\lambda_{Lm}|XC|\lambda_{Rm'}}}{2\sum_{m,m'}c_{Lm}c_{Rm'}\lambda_{m}^{|A|/2}\lambda_{m'}^{(N-|A|)/2}\braket{\lambda_{Lm}|XC|\lambda_{Rm'}}}.
\end{split}
\end{equation}
We take the thermodynamic limit $|A|,N\to\infty$ but fix the ratio $|A|/N\ll 1$ to be small, \eqnref{eq:dS1} will be dominated by the leading power ($m=m'=0$) and the sub-leading power ($m=1,m'=0$),
\begin{equation}\label{eq:dS2}
\begin{split}
\partial_{|A|}S_{\bar{\rho}_\infty}^{(2)}(A)&=-\frac{c_{L1}c_{R0}\lambda_{1}^{|A|/2}\lambda_{0}^{(N-|A|)/2}(\log\lambda_{1}-\log\lambda_{0})\braket{\lambda_{L1}|XC|\lambda_{R0}}}{2(c_{L0}c_{R0}\lambda_{0}^{N/2}\braket{\lambda_{L0}|XC|\lambda_{R0}}+c_{L1}c_{R0}\lambda_{1}^{|A|/2}\lambda_{0}^{(N-|A|)/2}\braket{\lambda_{L1}|XC|\lambda_{R0}})}\\
&=\frac{1}{2}\frac{\log(\lambda_0/\lambda_{1})}{\frac{c_{L0}\braket{\lambda_{L0}|XC|\lambda_{R0}}}{c_{L1}\braket{\lambda_{L1}|XC|\lambda_{R0}}}(\frac{\lambda_{0}}{\lambda_{1}})^{|A|/2}+1}.
\end{split}
\end{equation}
The behavior of $\partial_{|A|}S_{\bar{\rho}_\infty}^{(2)}(A)$ in the $|A|\to\infty$ limit crucially depends on whether or not $\frac{c_{L0}\braket{\lambda_{L0}|XC|\lambda_{R0}}}{c_{L1}\braket{\lambda_{L1}|XC|\lambda_{R0}}}$ vanishes or not. On general ground, $c_{L0}$ would not vanish, because it is a boundary condition that is chosen with some arbitrariness. So it all depends on the inner product $\braket{\lambda_{L0}|XC|\lambda_{R0}}$. If $\braket{\lambda_{L0}|XC|\lambda_{R0}}=0$, then $\partial_{|A|}S_{\bar{\rho}_\infty}^{(2)}(A)=\frac{1}{2}\log(\lambda_{0}/\lambda_{1})$. If $\braket{\lambda_{L0}|XC|\lambda_{R0}}\neq 0$, then as $|A|\to\infty$ the power $(\lambda_{0}/\lambda_{1})^{|A|/2}\to\infty$ diverges, hence $\partial_{|A|}S_{\bar{\rho}_\infty}^{(2)}(A)=0$. Therefore, the volume-law coefficient is determined by
\begin{equation}
f=\lim_{|A|\to\infty}\frac{\partial_{|A|}S_{\bar{\rho}_\infty}^{(2)}(A)}{\log d}=\left\{\begin{array}{ll}\frac{1}{2}\log_d(\lambda_0/\lambda_1) & \braket{\lambda_{L0}|XC|\lambda_{R0}}=0,\\
0 & \braket{\lambda_{L0}|XC|\lambda_{R0}}\neq 0.\end{array}\right.
\end{equation}
Using this formula, we calculated the volume-law coefficient for different measurement strength $p$ and different qudit dimension $d$, and the result is shown in \figref{fig:entropy}(c).

Now we explain our calculation of the measurement-induced entropy drop $\Delta S_x^{(2)}(A)$ and the qudit-environment mutual information $I_{\rho}({x}:A)$. Via the VUMPS algorithm, we have obtained the final entanglement feature state $\ket{W_{\rho_\infty}}$ as the leading eigenstate of $\hat{T}_{K}$ in \eqnref{eq:TKbrickwall}. Note that the last step of $\hat{T}_{K}$ is a layer of $\hat{T}_M$ (measurement). The state prepared by $\hat{T}_K$ is not quite what we want, because the qudits have been uniformly measured in the last step, then further probing the state with local measurement will double the effect of measurement and can not reflect the actual measurement-induced entropy drop right after the application of unitary gates. In order to prepare a ``fresh'' state right after the unitary layer, we apply an additional layer of unitary gate transfer matrix to the MPS state to construct the following entanglement feature state
\begin{equation}
\ket{W_\rho}=\cdots\dia{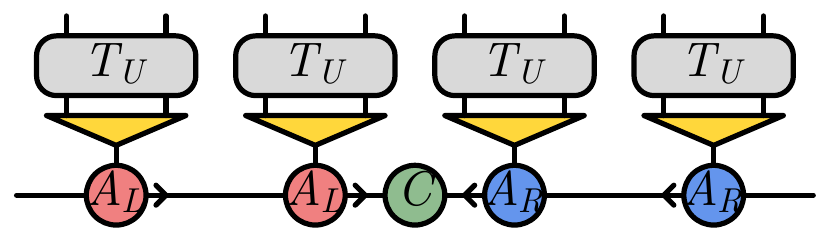}{42}{-6}\cdots.
\end{equation}
Now we can probe the system with a single-site measurement of strength $p$. This amounts to applying the transfer matrix $\hat{T}_{M_x}$ to $\ket{W_\rho}$ at site $x$,
\begin{equation}
\hat{T}_{M_x}\ket{W_\rho}=\cdots\dia{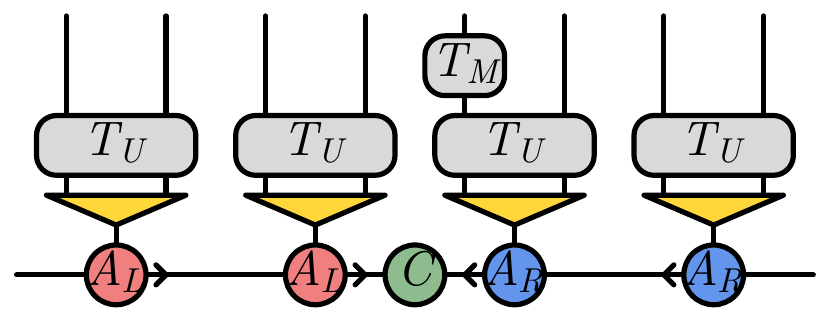}{56}{-6}\cdots.
\end{equation}
We can then compare the difference of entanglement entropies before and after the measurement in a region $A$ that encloses the site $x$,
\begin{equation}\label{eq:DeltaS_appB}
\Delta S_x^{(2)}(A)=-\log\frac{\braket{A|\hat{T}_{M_x}|W_\rho}}{\braket{\Uparrow|\hat{T}_{M_x}|W_\rho}}+\log\frac{\braket{A|W_\rho}}{\braket{\Uparrow|W_\rho}}.
\end{equation}
As explained in \appref{app:entropy}, this entropy drop is closely related to the qudit-environment mutual information $I_\rho(\{x\}:\bar{A})$, defined via
\begin{equation}\label{eq:expI_appB}
e^{I_\rho(\{x\}:\bar{A})}=\frac{\braket{A|X_x|W_\rho}\braket{\Uparrow|W_\rho}}{\braket{A|W_\rho}\braket{\Uparrow|X_x|W_\rho}},
\end{equation}
We will leave the explanations of \eqnref{eq:DeltaS_appB} and \eqnref{eq:expI_appB} to \appref{app:entropy} and focus on how to evaluate these quantities from the numerically obtained MPS in this appendix.

To help our calculation, we need to first define the following matrices
\begin{equation}
A_{R}^{T_M\uparrow}=\dia{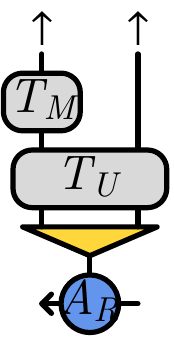}{60}{-5},\quad A_{R}^{T_M\downarrow}=\dia{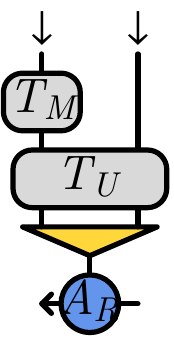}{60}{-5};\quad A_{R}^{\uparrow\downarrow}=\dia{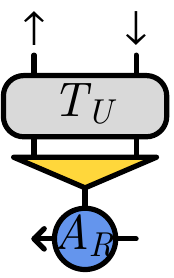}{48}{-5},\quad A_{R}^{\downarrow\uparrow}=\dia{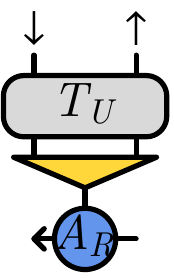}{48}{-5}.
\end{equation}
In fact, they are related by $\ZZ_2$ symmetry: $A_{R}^{T_M\downarrow}=XA_{R}^{T_M\uparrow}X$ and $A_{R}^{\downarrow\uparrow}=XA_{R}^{\uparrow\downarrow}X$. With these notations, we have
\begin{equation}\label{eq:inner_products}
\begin{split}
\braket{\Uparrow|W_\rho}&=\braket{\lambda_{L0}|C|\lambda_{R0}},\\
\braket{\Uparrow|\hat{T}_{M_x}|W_\rho}&=\lambda_0^{-1}\braket{\lambda_{L0}|CA_{R}^{T_M\uparrow}|\lambda_{R0}},\\
\braket{\Uparrow|X_x|W_\rho}&=\lambda_0^{-1}\braket{\lambda_{L0}|CA_{R}^{\downarrow\uparrow}|\lambda_{R0}},\\
\braket{A|W_\rho}&=\lambda_0^{-|A|/2}\braket{\lambda_{L0}|C(A_{R}^{\downarrow})^{|A|/2}|\lambda_{R0}},\\
\braket{A|\hat{T}_{M_x}|W_\rho}&=\lambda_0^{-|A|/2}\braket{\lambda_{L0}|(A_{L}^{\downarrow\intercal})^{(x-1)/2}CA_{R}^{T_M\downarrow}(A_{R}^{\downarrow})^{(|A|-x-1)/2}|\lambda_{R0}},\\
\braket{A|X_x|W_\rho}&=\lambda_0^{-|A|/2}\braket{\lambda_{L0}|(A_{L}^{\downarrow\intercal})^{(x-1)/2}CA_{R}^{\uparrow\downarrow}(A_{R}^{\downarrow})^{(|A|-x-1)/2}|\lambda_{R0}}.
\end{split}
\end{equation}
We have assumed that the region $A$ is embedded in a infinitely large system such that the boundary condition at the entanglement cuts are given by the eigenstates $\bra{\lambda_{L0}}$ and $\ket{\lambda_{R0}}$. Here $x$ is an integer labeling the position of the measurement site with respect to the entanglement cut. We assume that $x$ is odd to avoid more tedious discussion of the even-odd effect. For the purpose of studying the scaling behavior with respect to $x$, it is fine to probe only the odd sites. Given the expressions in \eqnref{eq:inner_products}, \eqnref{eq:DeltaS_appB} and \eqnref{eq:expI_appB} can be evaluated from the MPS tensors $A_L$, $C$ and $A_R$. Following this approach, we calculated $\Delta S_x^{(2)}(A)$ and $I_{\rho}({x}:A)$ at $d=2$ and $p=0.1$, using the MPS ansatz with bond dimension 64 (i.e.~$n=6$). The result is shown in \figref{fig:QEC}(c).

\section{Argument for Quantum Error Correcting Volume-Law State}
\label{app:QEC}
In this section, we give a self-consistent argument on the relation between the measurement-doped unitary circuit and quantum error correction. Our argument is directly motivated by toy examples, including the five-qubit code,  holographic codes, and more general stabilizer codes, but applies more generally without referring to any microscopic details. 

\subsection{Toy examples}

In this section, we describes two toy examples of error correcting states. Despite some differences, both constructions produce a sub-thermal volume-law state, the entanglement of which is robust against moderate amount of measurement.

\begin{rmk}[Five-qubit code]
The first toy example is constructed by taking a Page state
and encode each qubit into five qubits by the 5-qubit
QEC code, as depicted in \figref{fig:five qubit code}. The state
exhibits a volume-law entanglement with $f = 1/5$ on average and is stable against any measurement that acts on less than three qubits in every 5-qubit group. The QEC layer protects the quantum information of the Page state from being accessed by local measurements, hence the entanglement entropy can remain unchanged under measurements.

From this example, it is clear that such behavior is only possible in the sub-thermal volume-law state with $f < 1$, because it is those $(1-f)$ fraction of qubits that serve as the syndrome bits to enable QEC encoding of the Page state. Noticing that the code distance for the whole layer is only three, this state is not robust against probabilistic measurement. We need different blocks to have correlation, which inspires the next example.
\begin{figure}[t]
    \centering
    \includegraphics[width=0.25\textwidth]{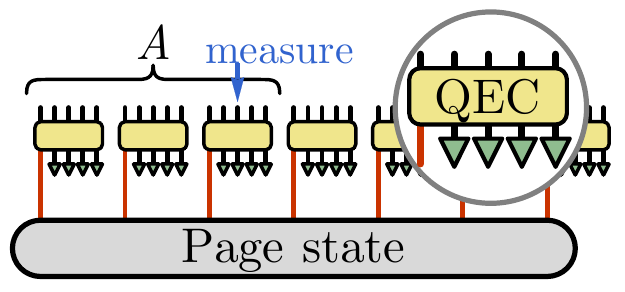}
    \caption{Five-qubit code toy model.}
    \label{fig:five qubit code}
\end{figure}
\end{rmk}

\begin{rmk}[Holographic code]
The second toy example is constructed by a random tensor network (RTN). Consider a system of $N$ qudits (each qudit is of Hilbert space dimension $d$), the Page state of these qudits admits a simple RTN representation as shown in \figref{fig:RTN}(a), where all physical legs are connected to a big random tensor $T_{\alpha_1\alpha_2\cdots\alpha_N}$ in the center. More precisely, the random tensor $T$ discribes the coefficient of the Page state when it is represented on a set of many-qudit basis states, 
\eq{\label{eq:Page}\ket{\Psi_\text{Page}}=\prod_{i=1}^N\sum_{\alpha_i=1}^d T_{\alpha_1\alpha_2\cdots\alpha_N}\ket{\alpha_1}\otimes\ket{\alpha_2}\cdots\otimes\ket{\alpha_N},}
where each tensor element in $T$ is randomly drawn from independent Gaussian distributions. Now we protect the Page state by one additional layer of matrix product operators (MPO) as shown in \figref{fig:RTN}(b).
\begin{equation}
    \label{eq:sub-Page}
    \ket{\Psi_\text{sub-Page}} = \prod_{i=1}^N \sum_{\beta_i=1}^{d_1}\hat{O}^{\beta_N}_{\beta_1}\hat{O}^{\beta_1}_{\beta_2} \cdots \hat{O}^{\beta_{N-1}}_{\beta_N} \ket{\Psi_\text{Page}},
\end{equation}
where $\hat{O}^{\beta_{i}}_{\beta_{i+1}}=\sum_{\alpha_i,\alpha'_i}\ket{\alpha_i}O^{\beta_{i}}_{\beta_{i+1}\alpha_i\alpha'_i}\bra{\alpha'_i}$ is the operator acting on the $i$th qudit as specified by 4-leg tensors of the form $O^{\beta}_{\beta'\alpha\alpha'}$ and is also randomly drawn.

\begin{figure}[htbp]
	\begin{center}
		\includegraphics[width=0.36\columnwidth]{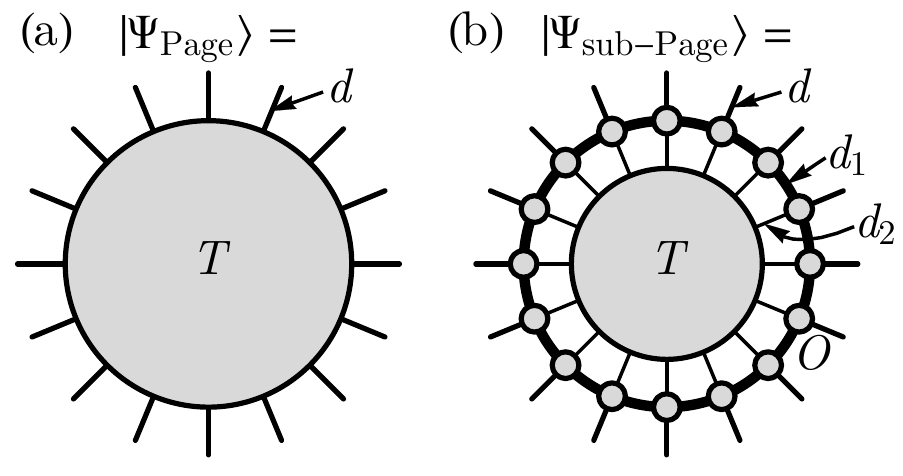}
		\caption{Random tensor network representations of (a) Page states and (b) sub-Page states. The qudit dimension is $d$. Bond dimensions of the matrix product operator are specified by $d_1$ and $d_2$, assuming $d_1>d>d_2$.}
		\label{fig:RTN}
	\end{center}
\end{figure}

As the tensors are random, the only relevant parameters of the MPO are its bond dimensions. As specified in \figref{fig:RTN}(b), we require the bond dimensions to satisfy the hierarchy $d_2<d$ (modeling introducing extra ancilla) and $d_1>d$ (modeling a few layers of local unitary circuit) The resulting state, called the sub-Page state in our discussion, is by construction sub-thermal and is robust against projective measurement.

Let us consider a subsystem, which is denoted by the red arrow in \figref{fig:effect of meaturement}, and the measurement on it.
For the Page state, any single measurement will disentangle the qudit from the rest of the system, and the entanglement cut will redirect itself to go through the projection operator, therefore the entropy drops by $\log d$, as shown in \figref{fig:effect of meaturement}~(a). However, for the sub-Page state, the entanglement cut will remain unchanged as shown in \figref{fig:effect of meaturement}~(b). If $d_1$ is sufficiently large such that we have $2 \log d_1 > \log d_2$, any attempt to cut through the projection operator will have more cost more, as shown in \figref{fig:effect of meaturement}~(c). In this case, the measurement does not result in any drop of the entanglement entropy. So the central page state can be protected from local measurements just by a layer of MPO with sufficiently large bond dimension $d_1 > \sqrt{d_2}$. We can treat this layer of MPO as a QEC encoding circuit (in fact, random tensors are asymptotically perfect, meaning that they automatically approximate QEC codes). This model works as long as $d_2 < d$, i.e. the volume law fraction $f = \log(d_2/d)< 1$. 

\begin{figure}[htbp]
\begin{center}
	\includegraphics[width=0.8\columnwidth]{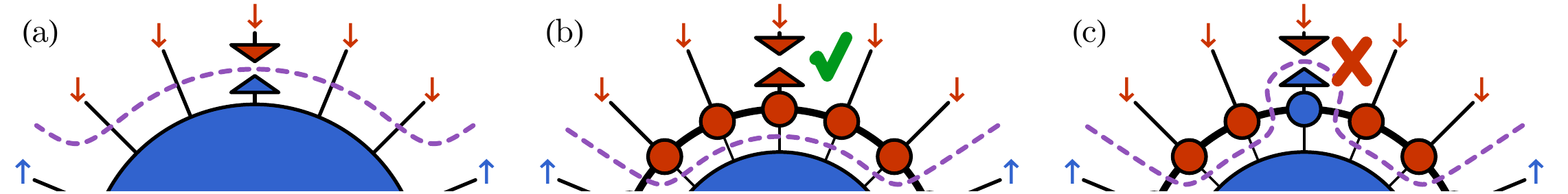}
	\caption{Response to the measurement for (a) the Page state and (b) the sub-Page state. The case shown in (c) is prohibited as long as $d_1>\sqrt{d_2}$.}
	\label{fig:effect of meaturement}
\end{center}
\end{figure}
\end{rmk}

\subsection{General argument}\label{app:qecc_argument}

In this section, we provide the general argument on why the final state can be understood by error correction.
As depicted in \figref{fig:QEC}~(a), we assume that the volume-law piece of the entanglement entropy of the final state completely comes from that of the input Page state. Namely, the QECC layer, regarded as a unitary transformation from the tensor product of the Page state and ancilla $\ket{\psi}\otimes\ket{0\cdots 0}$ to the final state $\ket{\chi}$, does \emph{not} increase the entanglement entropy of the original state by a volume-law amount. This locality constraint leads to the assumption that any large enough subsystem $A$ can have stabilizers that only have support on $A$.

This provides a natural mechanism to protect the entanglement from measurement, which is explained as follows. Let us call the final state $\ket{\chi}=T_{QECC}\ket{\psi}$ and consider the reduced density matrix for a large enough subsystem $A$. When measurements of of $t$ qubits happen in $A$, we can decompose the corresponding projection operator into a sum of Pauli strings as
\begin{equation}
	P_{i_1} P_{i_2} \cdots P_{i_t} = \mathcal{N} \left(\II + \sum_s c_s \scO_{s} \right),
\end{equation}
where $\mathcal{N}$ is a normalization factor and $\scO_s$ represents a Pauli string with weight equal or less than $t$. Accordingly, the purity of $\rho_A$ after the measurement can also be written as the following sum
\begin{equation}
		\Tr\rho_{A,after}^2 = \sum_{O_i,\tilde{O}_i}
	\begin{gathered}
	\includegraphics[width=2.5cm]{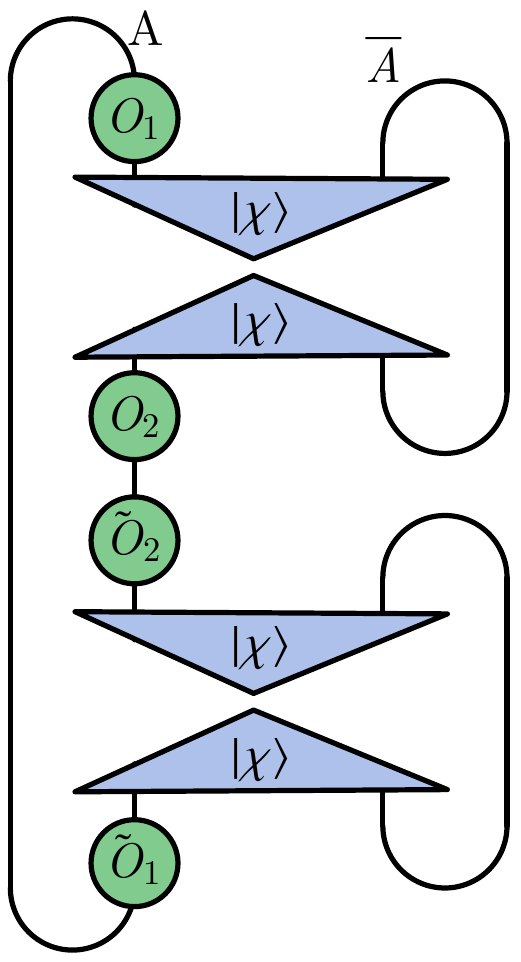}
	\end{gathered} = \sum_{\scO,\tilde{\scO}} 
	\begin{gathered}
	\includegraphics[width=2.5cm]{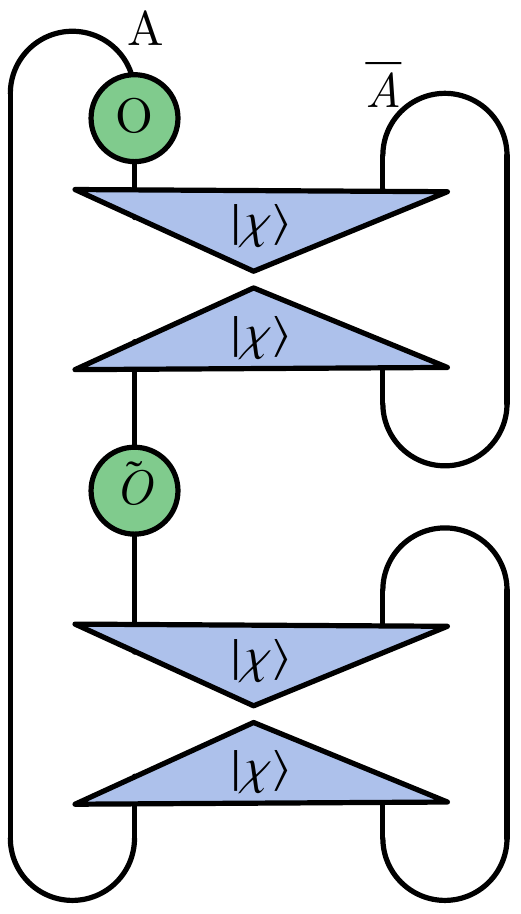}
	\end{gathered}\,,
\end{equation}
where we introduce $\scO = \scO_1\tilde{\scO_1}$ and $\tilde\scO = \scO_2\tilde{\scO_2}$. Now we assume $\scO$ and $\tilde\scO$ are detectable errors, which implies that they anti-commute with at least one stabilizer. When $\scO$ and $\tilde\scO$ are deep in the bulk of $A$, such stabilizers are fully supported in $A$ (they exist by assumption) and we can have for example
\begin{equation*}
		\begin{gathered}
		\includegraphics[width=2.5cm]{rhoApurity2}
		\end{gathered}
		=\begin{gathered}
		\includegraphics[width=2.5cm]{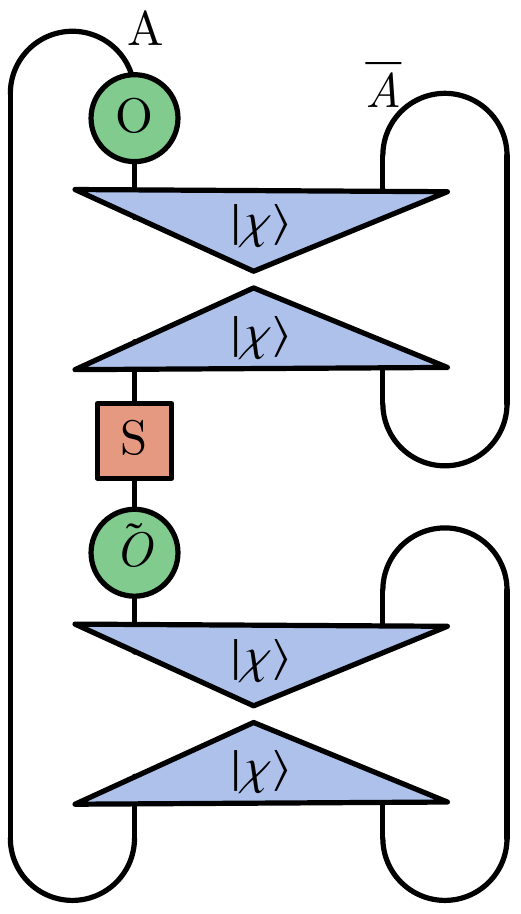}
		\end{gathered} 
		= - \begin{gathered}
		\includegraphics[width=2.5cm]{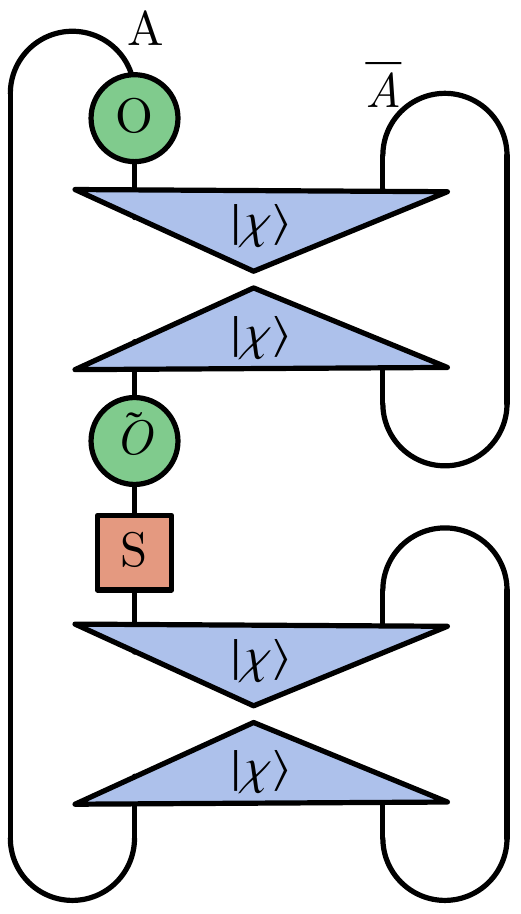}
		\end{gathered} 
	= -\begin{gathered}
		\includegraphics[width=2.5cm]{rhoApurity2}
	    \end{gathered}
\end{equation*}
for $\tilde{\scO}\neq \II$, which directly shows that such kind of terms vanishes. Similar calculations is true for $\scO$. The exceptions are when $\scO$ or $\tilde{\scO}$ are near the boundary of $A$ and the stabilizers may have support in both $A$ and $\bar A$. for $\tilde{\scO}\neq \II$, which directly shows that such kind of terms vanishes. Similar calculations is true for $\scO$. The exceptions are when $\scO$ or $\tilde{\scO}$ are near the boundary of $A$ and the stabilizers may have support in both $A$ and $\bar A$. It is easy to see that this argument still holds when the measurement is in $\bar A$ or both $A$ and $\bar A$.

More rigorously, we may consider a stabilizer QECC that can correct for any weight-$t$ Pauli error.  Consider the reduced density matrix for a subsystem $A$ in the codespace $\rho_{A}\equiv \Tr_{\bar{A}}(\Pi_{\mathrm{codespace}})$ where $\Pi_{\mathrm{codespace}}$ is the projector onto states in the codespace.  We now perform $m\le t$ single-qubit measurements in the Pauli basis, so that the new reduced density matrix is given by
$\sigma_{A} \propto \Pi_{A}\rho_{A}\Pi_{A}$,  
where $\Pi_{A}$ is a product of $m$ single-qubit projectors in the Pauli basis.  We may expand $\Pi_{A}$ as 
\begin{align}
\Pi_{A} = \frac{1}{2^{m}}\left[1 + \sum_{j=1}^{2^{m}-1}\mathcal{E}_{j}\right].
\end{align}
where $\{\mathcal{E}_{j}\}$ are Pauli operators.  
As we prove in the following section, the second R\'{e}nyi entropy $S^{(2)}(\rho_{A}) \equiv -\log_{2}\Tr(\rho_{A}^{2})$ is related to the entanglement for the same subsystem, after performing these $m$ single-qubit measurements in the Pauli basis, as
\begin{align}\label{eq:ent_reduction}
S^{(2)}(\sigma_{A}) = S^{(2)}(\rho_{A}) - \log_{2}[1 + n_{{A}}]
\end{align}
where $n_{A}$ is the number of Pauli operators in the set $\{\mathcal{E}_{j}\}$ that have syndromes which cannot be determined by performing measurements of stabilizers that are exclusively within the $A$ subsystem.  Assuming that the stabilizers have a finite average size, the quantity $n_{A}$ will scale exponentially in the number of measurements that are performed near the boundary of region $A$, so that the right-hand side of Eq. (\ref{eq:ent_reduction}) will only provide an area-law correction to the R\'{e}nyi entropy.  

If we roughly use $pN$ as the number of measurement in each round, then the code distance has to be larger than $pN$. Notice that the length of the stabilizers is not tightly constrained by the code distance. Therefore, although the code distance is macroscopic, the stabilizers can still have a microscopic length for our argument to work.

For this mechanism to continue work, all the errors have to be corrected by the next layer of unitary evolution, namely all the Pauli strings in the measurements are correctable errors. Therefore, the code distance has to be larger than $2pN$. If we assume the code distance is exactly $2pN$ (as well as the code being non-degenerate), then we can derive the Hamming bound shown in the main text. 

 Eq. (\ref{eq:ent_reduction}) can also be used to argue for the power-law decrease in the entanglement entropy when performing a measurement a distance $x$ from the boundary of a subsystem, in the sub-thermal volume-law phase that is obtained for Clifford dynamics with measurements in the Pauli basis. This is because Eq. (\ref{eq:ent_reduction}) also holds for any stabilizer state in which a single-qubit measurement has no overlap with the stabilizer group.  Consider a semi-infinite region $A$.  For the sub-thermal volume law state generated by random Clifford dynamics with measurements, let $p(x)$ be the probability that a single-qubit measurement, performed a distance $x$ from the boundary of $A$ commutes with all operators that stabilize the state, and that lie entirely within in the $A$ subsystem.  From Eq. (\ref{eq:ent_reduction}), the average entanglement entropy drop after this measurement is exactly
\begin{align}
\overline{\Delta S(x)} &\equiv [1-p(x)]\log_{2}(1) + \log_{2}(2)p(x) \nonumber\\
&= p(x).
\end{align}
Therefore, if $p(x)$ falls faster than $1/x$, then the the entanglement drop after performing a finite density of measurements will be a \emph{constant}.  

We estimate $p(x)$ using the known stabilizer length distribution $P(\ell)$ in random Clifford circuits, with measurements in the Pauli basis \cite{Li2019METHQC}. In the volume-law phase, it is known \cite{Li2019METHQC} that $P(\ell) = \alpha(p)\ell^{-2} + s(p)\delta(\ell - (L/2))$ in a system with size $L$.  We now consider a region $A$ defined by the interval $[1,|A|]$, and we perform a measurement at a position $x$ such that $1 \ll x\ll |A|$ where we perform a single measurement. The number of stabilizers that are contained entirely within $A$, and that have ``crossed" the position $x$, i.e. that have their left endpoint $y_{L}<x$ and their right endpoint $y_{R}>x$ is
\begin{align}
    N(x) \equiv \int_{0}^{x}\int_{x}^{|A|}dy_{L}\, dy_{R}\,P(|y_{L}-y_{R}|) = \alpha\ln(x) + O(x/|A|)
\end{align}
The probability that all of these stabilizers commute with the single-qubit measurement is exponentially small in the number of stabilizers, which gives an estimate of $p(x) \sim e^{-N(x)} = x^{-\alpha}$.  This then gives the power-law decay $\overline{\Delta S(x)} \sim x^{-\alpha}$ for the entanglement with the distance that the measurement is performed, from the boundary. The precise exponent for this power-law behavior cannot be determined without knowing more detailed properties of the stabilizers.  For example, if we assume that the stabilizers drawn from the distribution $P(\ell)$ have equal probability of acting as a Pauli $X$, $Y$, $Z$, or the identity $I$ at site $x$, then the probability $p(x) = 2^{-N(x)} = x^{-\alpha \ln(2)}$.

\subsection{Proof of Eq. (\ref{eq:ent_reduction})}
Let the tensor $T$ be the encoding of a state on $k$ qubits into a state on $N$ qubits with a stabilizer quantum error-correcting code (QECC).  We assume that this encoding is a valid quantum error-correcting code (QECC) with code distance $d$; the code can then correct for any Pauli error of weight $t \le \lfloor (d-1)/2 \rfloor$. Since $T:\mathbb{C}^{2^{k}}\rightarrow \mathbb{C}^{2^{N}}$ is a valid encoding map for a QECC, it is  isometric $T^{\dagger}T = 1_{2^{k}\times2^{k}}$. 

We now consider the density matrix 
\begin{align}
\rho\equiv TT^{\dagger}
\end{align}
which is a projector onto the codespace of the QECC.  We further bipartition the $N$ spins into an $A$ subsystem, and its complement $\bar{A}$, and define the reduced density matrix $\rho_{A} \equiv \Tr_{\bar{A}}(TT^{\dagger})$.  If $G$ is the Pauli stabilizer group for the QECC, then the reduced density matrix may be equivalently written as
\begin{align}\label{eq:rdm}
\rho_{A} = \frac{1}{D_{A}}\sum_{g\in G_{A}}g
\end{align}
where $G_{A}$ is the subgroup of $G$, consisting of elements of the stabilizer group that act as the identity operator on $\bar{A}$, and $D_{A}$ is the Hilbert space dimension of the $A$ subsystem.
  
Now, let $\Pi_{A}$ be a product of single-qubit projection operators in the Pauli basis, on $m \le t$ spins in the $A$ subsystem.   We may expand $\Pi_{A}$ as a sum of Pauli operators as
\begin{align}
\Pi_{A} = \frac{1}{2^{m}}\left[1 + \sum_{j=1}^{2^{m}-1}\mathcal{E}_{j}\right].
\end{align}
We refer to the Pauli operators $\{\mathcal{E}\}$ appearing in this expansion as ``errors".  Since $m\le t$, each of these  errors are correctable, and we observe that 
\begin{align}\label{eq:qecc_condition}
\Tr( \mathcal{E}_{j}\rho)=0 \hspace{.2in} \Tr( \mathcal{E}_{i}\mathcal{E}_{j}\rho)=0 \hspace{.1in} (i\ne j)
\end{align}
  As a result, the reduced density matrix for the state, after performing these measurements is 
\begin{align}
\sigma_{A}   \equiv \frac{\Pi_{A}\rho_{A}\Pi_{A}}{\braket{\Phi|\Pi_{A}|\Phi}} = 2^{m}\Pi_{A}\rho_{A}\Pi_{A}
\end{align}
The purity of $\sigma_{A}$ may be expanded as 
\begin{align}
\Tr(\sigma_{A}^{2}) = \Tr(\rho_{A}^{2}) + 2\sum_{j}\Tr(\mathcal{E}_{j}\rho_{A}) + \sum_{i,j}\Tr(\mathcal{E}_{i}\rho_{A}\mathcal{E}_{j}\rho_{A})\nonumber
\end{align}
We observe that $\Tr(\mathcal{E}_{j}\rho_{A}) = 0$ due to Eq. (\ref{eq:qecc_condition}).  

We evaluate the final term as follows.  First, we observe that $\Tr(\mathcal{E}_{i}\rho_{A}\mathcal{E}_{j}\rho_{A}) = 0$ if either $\mathcal{E}_{i}$ or $\mathcal{E}_{j}$ is an error with an \emph{localizable syndrome}, i.e. an error that can be detected via syndrome measurements that act exclusively in the $A$ subsystem. Let $\mathcal{E}_{i}$ be a localizable error; then there is an element $h\in G_{A}$, such that $\{h, \mathcal{E}_{i}\} = 0$.  As a result, 
\begin{align}
&\Tr(\mathcal{E}_{i}\rho_{A}\mathcal{E}_{j}\rho_{A}) = \Tr(\mathcal{E}_{i}h \rho_{A}\mathcal{E}_{j}\rho_{A})\nonumber\\
&= -\Tr(h\mathcal{E}_{i} \rho_{A}\mathcal{E}_{j}\rho_{A}) = -\Tr(\mathcal{E}_{i}\rho_{A}\mathcal{E}_{j}\rho_{A})
\end{align} 
so that $\Tr(\mathcal{E}_{i}\rho_{A}\mathcal{E}_{j}\rho_{A}) = 0$.  If both $\mathcal{E}_{i}$ and $\mathcal{E}_{j}$ cannot be localized, then both errors commute with the stabilizer subgroup $G_{A}$, and 
 \begin{align}
\Tr(\mathcal{E}_{i}\rho_{A}\mathcal{E}_{j}\rho_{A}) = \Tr(\mathcal{E}_{i}\mathcal{E}_{j}\rho_{A}^{2}) = \delta_{ij}\Tr(\rho_{A}^{2})
\end{align}
In the last line, we have again used Eq. (\ref{eq:qecc_condition}).  Therefore, we conclude that the second R\'{e}nyi entropy $S^{(2)}(\sigma_{A}) \equiv -\log_{2}\Tr(\sigma_{A}^{2})$ after the measurements is 
\begin{align}
S^{(2)}(\sigma_{A}) = S^{(2)}(\rho_{A}) - \log_{2}[1 + n_{{A}}]
\end{align}
where $n_{{A}}$ is the number of errors in $\{\mathcal{E}_{i}\}$ whose syndromes cannot be localized to the $A$ subsystem.  

\section{Entropy Drop and Qudit-Environment Information}
\label{app:entropy}
We propose the qudit-environment mutual information $I_\rho(\{x\}:\bar{A})=S_{\bar{\rho}}^{(2)}(\{x\})+S_{\bar{\rho}}^{(2)}(\bar{A})-S_{\bar{\rho}}^{(2)}(\{x\}\cup \bar{A})$ as a measure of the QEC capacity of the sub-thermal volume-law state. Note that the entanglement entropies are evaluated with respect to the normalized density matrix $\bar{\rho}=\rho/\Tr\rho$, such that
\begin{equation}\label{eq:expS}
e^{-S_{\bar{\rho}}^{(2)}(A)}=W_{\bar{\rho}}(A)=\frac{W_{\rho}(A)}{W_{\rho}(\emptyset)}=\frac{\braket{A|W_\rho}}{\braket{\Uparrow|W_\rho}},
\end{equation}
where $\ket{A}=\prod_{i\in A}X_i\ket{\Uparrow}$ is the Ising basis state for region $A$ (i.e.~$\sigma_i=\downarrow$ if $i\in A$ and $\sigma_i=\uparrow$ if $i\in\bar{A}$). Using \eqnref{eq:expS}, it can be shown that
\begin{equation}\label{eq:expI_app}
\begin{split}
e^{I_\rho(\{x\}:\bar{A})}&=e^{S_{\bar{\rho}}^{(2)}(\{x\})+S_{\bar{\rho}}^{(2)}(\bar{A})-S_{\bar{\rho}}^{(2)}(\{x\}\cup \bar{A})}\\
&=\frac{e^{-S_{\bar{\rho}}^{(2)}(\{x\}\cup \bar{A})}}{e^{-S_{\bar{\rho}}^{(2)}(\{x\})}e^{-S_{\bar{\rho}}^{(2)}(\bar{A})}}\\
&=\frac{\frac{\braket{A|X_x|W_\rho}}{\braket{\Uparrow|W_\rho}}}{\frac{\braket{\Uparrow|X_x|W_\rho}}{\braket{\Uparrow|W_\rho}}\frac{\braket{A|W_\rho}}{\braket{\Uparrow|W_\rho}}}\\
&=\frac{\braket{A|X_x|W_\rho}\braket{\Uparrow|W_\rho}}{\braket{A|W_\rho}\braket{\Uparrow|X_x|W_\rho}},
\end{split}
\end{equation}
which explains \eqnref{eq:expI}.

Consider making a measurement at position $x$ in region $A$. Suppose the measurement is described by the operator $M_x$, its effect on the entanglement feature is implemented by acting the corresponding transfer matrix $\hat{T}_{M_x}$ to the entanglement feature state $\ket{W_\rho}\to\hat{T}_{M_x}\ket{W_\rho}$. According to \eqnref{eq:expS}, the entanglement entropy of region $A$ after the measurement is given by
\begin{equation}
S_x^{(2)}(A)=-\log\frac{\braket{A|\hat{T}_{M_x}|W_\rho}}{\braket{\Uparrow|\hat{T}_{M_x}|W_\rho}},
\end{equation}
where the denominator $\braket{\Uparrow|\hat{T}_{M_x}|W_\rho}$ provides the appropriate normalization for the entanglement feature state. Therefore the entropy drop after measurement should be defined as
\begin{equation}\label{eq:DeltaS_app}
\Delta S_x^{(2)}(A)=-\log\frac{\braket{A|\hat{T}_{M_x}|W_\rho}}{\braket{\Uparrow|\hat{T}_{M_x}|W_\rho}}+\log\frac{\braket{A|W_\rho}}{\braket{\Uparrow|W_\rho}},
\end{equation}
which is the definition given in \eqnref{eq:DeltaS}.

To derive the relation between the measurement-induced entropy drop $\Delta S^{(2)}_x(A)$ and the qudit-environment mutual information $I_\rho(\{x\}:\bar{A})$, we start with the definition in \eqnref{eq:DeltaS_app},
\begin{equation}
\begin{split}
    \Delta S_x^{(2)}(A)&\equiv-\log\frac{\braket{A|\hat{T}_{M_x}|W_\rho}}{\braket{\Uparrow|\hat{T}_{M_x}|W_\rho}}+\log\frac{\braket{A|W_\rho}}{\braket{\Uparrow|W_\rho}}\\
    &=-\log\frac{\braket{A|1-\frac{p}{d+1}+\frac{pd}{d+1}X_x|W_\rho}}{\braket{\Uparrow|1-\frac{p}{d+1}+\frac{pd}{d+1}X_x|W_\rho}}+\log\frac{\braket{A|W_\rho}}{\braket{\Uparrow|W_\rho}}\\
    &=-\log\frac{\braket{A|1+\frac{pd}{d+1-p}X_x|W_\rho}}{\braket{\Uparrow|1+\frac{pd}{d+1-p}X_x|W_\rho}}+\log\frac{\braket{A|W_\rho}}{\braket{\Uparrow|W_\rho}}\\
    &=-\log\Big(1+\frac{pd}{d+1-p}\frac{\braket{A|X_x|W_\rho}}{\braket{A|W_\rho}}\Big)+
    \log\Big(1+\frac{pd}{d+1-p}\frac{\braket{\Uparrow|X_x|W_\rho}}{\braket{\Uparrow|W_\rho}}\Big),
\end{split}
\end{equation}
where we have used inserted the definition of $\hat{T}_{M_x}$ in \eqnref{eq:TM}. Assuming $p$ is small in the weak measurement limit, we expand $\Delta S_x^{(2)}(A)$ in power series of $p$,
\begin{equation}
\begin{split}
    \Delta S_x^{(2)}(A)&=-\frac{pd}{d+1}\Big(\frac{\braket{A|X_x|W_\rho}}{\braket{A|W_\rho}}-\frac{\braket{\Uparrow|X_x|W_\rho}}{\braket{\Uparrow|W_\rho}}\Big)+\scO(p^2)\\
    &=-\frac{pd}{d+1}\frac{\braket{\Uparrow|X_x|W_\rho}}{\braket{\Uparrow|W_\rho}}\Big(\frac{\braket{A|X_x|W_\rho}\braket{\Uparrow|W_\rho}}{\braket{A|W_\rho}\braket{\Uparrow|X_x|W_\rho}}-1\Big)+\scO(p^2)\\
    &=-\frac{pd}{d+1}W_{\bar{\rho}}(\{x\})\big(e^{I_{\rho}(\{x\}:\bar{A})}-1\big)+\scO(p^2),
\end{split}
\end{equation}
where $\frac{\braket{\Uparrow|X_x|W_\rho}}{\braket{\Uparrow|W_\rho}}=W_{\bar{\rho}}(\{x\})=e^{-S_{\bar{\rho}}^{(2)}(\{x\})}$ is the single-qudit purity (at position $x$). We have used \eqnref{eq:expI_app} to introduce the exponentiated mutual information $e^{I_\rho(\{x\}:\bar{A})}$. In the volume-law phase, a rough estimate is $S_{\bar{\rho}}^{(2)}(\{x\})=f\log d$, hence $W_{\bar{\rho}}(\{x\})=d^{-f}$, therefore
\begin{equation}
    \Delta S_x^{(2)}(A)=-p\big(e^{I_{\rho}(\{x\}:\bar{A})}-1\big)\frac{d^{1-f}}{d+1}+\scO(p^2),
\end{equation}
which justifies the relation of \eqnref{eq:DeltaS}. If $x$ is deep in region $A$, the mutual information $I_\rho(\{x\}:\bar{A})$ is expected to be small. In the limit of $I_\rho(\{x\}:\bar{A})\to0$, the entropy drop $\Delta S_x^{(2)}(A)$ directly proportional to the mutual information $I_\rho(\{x\}:\bar{A})$,
\begin{equation}
	\Delta S_x^{(2)}(A)\simeq -p\frac{d^{1-f}}{d+1} I_{\rho}(\{x\}:\bar{A}).
\end{equation}
The more the qudit $x$ can inform about the complement region $\bar{A}$, the more entropy drop will be produced by measuring it.

\section{Fermionic Gaussian State Approximations}
\label{app:fermion}

In this section, we give an analytical calculation of the exponent $3/2$ using the free fermion approximation. We start by analyzing the transfer matrices $\hat{T}_{U_{ij}}$ and $\hat{T}_{M_i}$ in \eqnref{eq:TUTM}. Recall that the entanglement is mapped to Ising spin correlation in the entanglement feature formulation. The unitary gate entangles the nearby sites together, hence $\hat{T}_{U_{ij}}$ generally promotes the ferromagnetic correlations between neighboring Ising spins. The on-site measurement disentangles the qudit from its environment, hence $\hat{T}_{M_i}$ generally breaks the Ising correlation and disorder the spin. Therefore, it is reasonable to approximate $\hat{T}_{U_{ij}}\simeq e^{JZ_iZ_j}$ and $\hat{T}_{M_i}\simeq e^{hX_i}$ by the imaginary time evolution of Ising coupling and transverse field terms respectively. This approximation allows us to simplify the entanglement dynamics to an imaginary time Floquet problem of quantum Ising model, which can then be mapped to a free fermion Floquet problem and solved analytically. In this way, we can obtain the exponent $3/2$ analytically. In the simplified model, the one-step transfer matrix for the entanglement feature state reads
\begin{equation}
    \label{eqn:simplified Tstep}
    \hat T_{\text{step}} = \prod_{i=1}^{N} e^{J Z_i Z_{i+1}} \prod_{i=1}^{N} e^{h X_i}\,,
\end{equation}
where $N$ is the system size assuming the periodic boundary condition. We keep a finite $N$ to regulate the calculation and take the thermodynamic limit  ($N\rightarrow\infty$) in the end. The relative ordering between $\hat{T}_{U_{ij}}$ and $\hat{T}_{M_i}$ does not change results qualitatively. Here, we put $\hat{T}_{U_{ij}}$ on the left side of $\hat{T}_{M_i}$, as contrary to the ordering in \eqnref{eq:Tstep}, such that it prepares a final state suitable for studying the measurement effect. 
\eqnref{eqn:simplified Tstep} imitates a Trotterized $(1+1)$D transverse field Ising model in the imaginary-time and can be exactly solvable by a Jordan-Wigner transformation \begin{equation}\label{eq:JW}
	\chi_{2j-1} = \prod_{1\le i<j} X_i Z_j\,,\quad
	\chi_{2j} = \prod_{1\le i < j} X_i Y_j\,,\quad
	j=1,2,\cdots,N\,,
\end{equation}
with $\chi_{1} = Z_1$ and $\chi_2 = Y_1$. The $\ZZ_2$ symmetry operator $\prod_iX_i$ of the Ising spins is also the fermion parity operator of the Jordan-Wigner fermions. Since the Ising model is restricted to the $\ZZ_2$ even sector, the fermions are also in the $\ZZ^F_2$ even sector with the anti-periodic boundary condition. The transfer matrix rewritten in terms of fermions is
\begin{equation}
	\hat{T}_\text{step} = \exp\left( \ii J \sum_{j=1}^N \chi_{2j}\chi_{2j+1} \right)\exp\left( \ii h \sum_{j=1}^N \chi_{2j-1}\chi_{2j} \right)\,,\quad \chi_{2N+1} = -\chi_1\,.
\end{equation}
We can diagonalize the transfer matrix using the fermion formalism in the momentum space. We first define the momentum-space fermion operators $c_{k,A}$ and $c_{k,B}$ with two sites (labeled by $A$ and $B$) per unit cell,
\begin{equation}
	\chi_{2j-1} = \frac{1}{\sqrt{N}} \sum_{k} e^{\ii jk} c_{k,A}\,,\quad
	\chi_{2j} = \frac{1}{\sqrt{N}} \sum_{k} e^{\ii jk} c_{k,B}\,.
\end{equation}
The momentum takes the values in $k\in[-\pi,\pi)$ with $k = \frac{2\pi}{N}\left(s+\frac{1}{2} \right), s\in\ZZ$. For simplicity, $N$ is fixed to be an even number in order to avoid the $k=\pi$ mode. Notice that $c_{k,A/B}$ are complex fermions with the $k<0$ modes being related to the $k>0$ modes by $c_{-k,A} = c_{k,A}^\dag$ and $c_{-k,B} = c_{k,B}^\dag$, so that the $k<0$ modes can be excluded to avoid double counting.
As a result, the transfer matrix cast in the momentum space can be factorized into a product of each momentum mode
\begin{equation}\label{eq:T(k)}
	\hat{T}_{\text{step}} = \prod_{k>0} \exp\left(-c_k^\dag h_k^J c_k\right) \exp\left(-c_k^\dag h_k^h c_k\right)=\prod_{k>0}\hat{T}(k)\,,
\end{equation}
where we have defined $c_k = \binom{c_{k,A}}{c_{k,B}}$ and
\eqs{
	\hat{T}(k)&=\exp\left(-c_k^\dag h_k^J c_k\right) \exp\left(-c_k^\dag h_k^h c_k\right),\\
	h_k^J & = J\left(\sin k \sigma^x -\cos k \sigma^y\right),\\
	h_k^h &= h \sigma^y.}
Different momentum modes can be diagonalized separately. Let us introduce
\begin{equation}
	\label{eqn:a b definition}
	a_k = \cosh J \cosh h - e^{\ii k} \sinh J \sinh h\,,\quad
	b_k = \cosh J \sinh h - e^{\ii k} \sinh J \cosh h\,.
\end{equation}
Then the leading eigenvalue of $\hat{T}(k)$ is given by
\eq{\lambda_{k,+}=\Re a_k+\sqrt{|b_k|^2-(\Im a_k)^2},}
and the corresponding leading eigenstate can be written as
\begin{equation}
	\label{eqn:leading state}
	\begin{aligned}
	\ket{W_{\rho_\infty}} \propto \prod_{k>0} \ket{\lambda_{k,+}} = \prod_{k>0} \left( A_k c_{k,A}^\dag + B_k c_{k,B}^\dag \right) \ket{\text{vac}}\,. \\
	A_k = \ii b_k^*\,, B_k = \ii \Im a_k + \sqrt{|b_k|^2 - (\Im a_k)^2}\,,
	\end{aligned}
\end{equation}
with $\ket{\text{vac}}$ being the vacuum state of $c_{k,A/B}$. One can check that $\ket{W_{\rho_\infty}}$ always has an even fermion parity and thus is indeed a legitimate entanglement feature state (respecting the Ising symmetry $\ZZ_2$ in the spin language). It will be useful mention that in the limit of $h=0$, the state $\ket{W_{\rho_\infty}}$ reduces to
\eq{\label{eq:FM cat state}\ket{W_{h=0}}\propto\prod_{k>0}(-\ii e^{-\ii k}c_{k,A}^\dagger+c_{k,B}^\dagger)\ket{\text{vac}},}
which corresponds to $\ket{W_{h=0}} = \ket{\Uparrow} + \ket{\Downarrow}$ in the spin language, because the transfer matrix contains only the Ising coupling term $\prod_{i}e^{J Z_i Z_{i+1}}$ in this limit, whose leading eigenstate is the ferromagnetic cat state.

Having found the leading eigenstate $\ket{W_{\rho_\infty}}$ of the transfer matrix $\hat{T}_\text{step}$, we can evaluate the entanglement feature in any region $A$ by
\eq{\label{eq:W(A)0}W_{\bar{\rho}_\infty}(A) =\frac{\braket{A|W_{\rho_\infty}}}{\braket{\Uparrow|W_{\rho_\infty}}},}
where $\ket{A} = \prod_{i\in A}X_i\ket{\Uparrow}$ encodes the region $A$ and $\ket{\Uparrow}$ is the all-up state in the Ising language. To proceed, we notice that the $\ZZ_2$ symmetry of the state $\ket{W_{\rho_\infty}}$ allows us to replace $\ket{\Uparrow}$ by its $\ZZ_2$ symmetric form $\ket{\Uparrow} + \ket{\Downarrow}=\ket{W_{h=0}}$ without affecting the result. This amounts to the following replacements
\eqs{\label{eq:Z2 replace}\ket{\Uparrow}&\to\ket{W_{h=0}},\\
\ket{A}=\prod_{i\in A}X_i\ket{\Uparrow}&\to\prod_{i\in A}X_i\ket{W_{h=0}}=Z_{i_0}\left(\prod_{i_0<i<i_1}X_i\right)Z_{i_1}\ket{W_{h=0}}=\ii\chi_{2i_0}\chi_{2i_1-1}\ket{W_{h=0}},}
where we have assumed the region $A$ to be a single segment strictly between sites $i_0$ and $i_1$ (assuming $i_1> i_0$, such that $|A|=i_1-i_0-1$ counts the size of $A$). In the above derivation, we are free to insert the $Z_{i_0}Z_{i_1}$ operator because the state $\ket{W_{h=0}} = \ket{\Uparrow} + \ket{\Downarrow}$ has fully correlated that $Z_{i_0}Z_{i_1}\ket{W_{h=0}}=\ket{W_{h=0}}$. Then the string operator dressed by the $Z$ operators can be translated to the fermion bilinear operator following \eqnref{eq:JW}. Plugging \eqnref{eq:Z2 replace} into \eqnref{eq:W(A)0}, we arrive at
\begin{equation}\label{eq:W(A)}
	W_{\bar{\rho}_\infty}(A) = \frac{\braket{W_{h=0} | \ii \chi_{2i_0} \chi_{2i_1-1} | W_{\rho_\infty}}}{\braket{W_{h=0}| W_{\rho_\infty}}}\,.
\end{equation}
which explains \eqnref{eq:Winf_def} by taking $i_0=0$ and $i_1=|A|+1$. One can also choose to insert any of the four combination of $Z_{i_0}/Z_{i_0-1}$ and $Z_{i_1}/Z_{i_1+1}$ and they yield different fermion operators by by construction give the same result. This gauge choice comes from the fact that $\ket{W_{h=0}}$ appears on the left of the correlator.
Note that we denote the numerator of \eqnref{eq:W(A)} by $W_{\rho_\infty}(A)=\braket{W_{h=0} |\ii \chi_{2i_0} \chi_{2i_1-1} | W_{\rho_\infty}}$, which is the unnormalized entanglement feature.

\begin{figure}[t]
	\centering

	\includegraphics[width=0.75\textwidth]{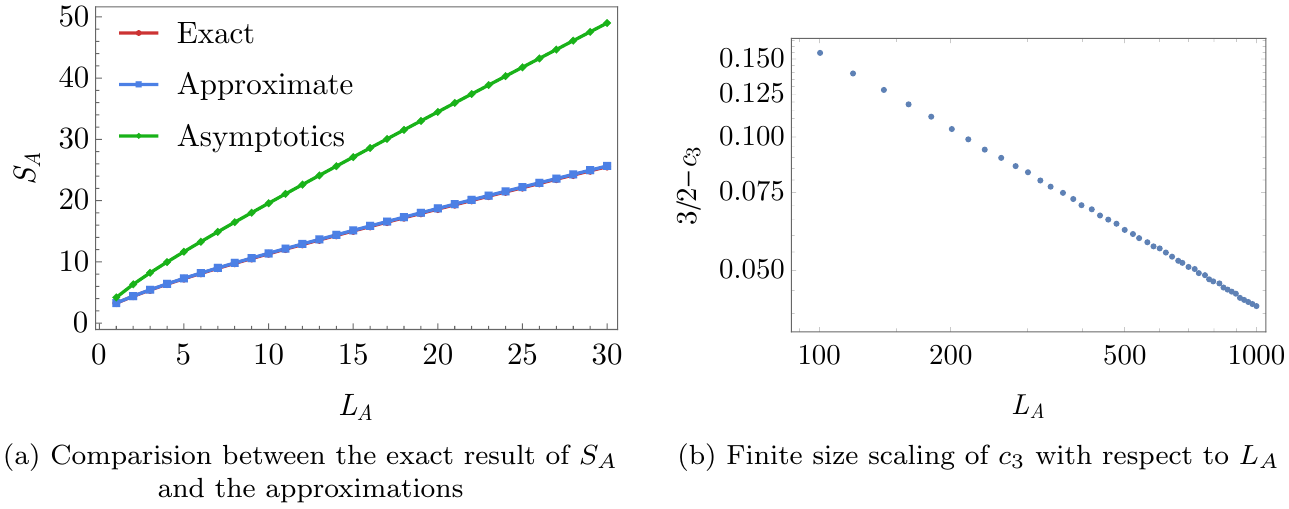}
	\caption{(a)Comparison between the exact result of $S_A$, the approximated result using contour integral and the asymptotic expansion of that. The parameter for the plot is $J=1,h=2/5$. The constant $C$ is chosen as $\frac{\sqrt{(z_3-\bar z)(z_4 - \bar z)}}{1-\bar{z}^2}$ with $\bar z = \frac{z_1+z_2}{2}$ being the middle point of the integration range. (b) We generate the entanglement data using \eqnref{eqn:Wsigma contour}, fit it by $c_1+c_2|A|+c_3\log |A|$ and study $c_3$'s dependence on $|A|$. It can be seen that $c_3$ exponentially converges to $3/2$ as predicted by \eqnref{eqn:Wsigma asymptotics}. }
	\label{fig:entropy contour}
\end{figure}

Given the fermion Gaussian states $\ket{W_{\rho_\infty}}$ in \eqnref{eqn:leading state} and $\ket{W_{h=0}}$ in \eqnref{eq:FM cat state}, it is straightforward to evaluate $W_{\bar{\rho}_\infty}$ in \eqnref{eq:W(A)}, and in the thermodynamic limit, the result reads
\begin{equation}
	\label{eqn:Wsigma of A}
	W_{\bar{\rho}_\infty}(A) = \frac{\ii}{2\pi} \int_{-\pi}^{\pi} \dd k R_k e^{\ii k |A|}\,,\,\,
	R_k = \frac{\Im(\tilde{A}_k B_k^*)}{|\tilde{A}_k|^2 + \Re(\tilde{A}_k B_k^*)}\,,
\end{equation}
where $\tilde{A}_k=\ii e^{\ii k}A_k=-e^{\ii k}b_k^*$. Let us compute the integral using the contour integral method. We rewrite $R_k$ as a function of $z=e^{\ii k}$ as follows
\begin{equation}
	\label{eqn:Rz definition}
	\begin{gathered}
	R(z) = \ii \frac{P_2(z) - \sinh J \sinh h \sqrt{P_4(z)}}{e^J \sinh h(z^2-1)}\,, \\ 
	P_2(z) = 2\sinh J \cosh h z - \cosh J \sinh h (z^2 + 1)\,,\\
	P_4(z) = (z - z_1)(z - z_2)(z - z_3)(z - z_4)\,,0<z_1<z_2<1,z_1 z_4 = z_2 z_3 = 1\,.
	\end{gathered}
\end{equation}
$P_4(z)$ is a four-th order polynomial with two of its zeros sitting inside the unit circle and the other two sitting outside.
When writing $\sqrt{P_4(z)}$, we implicitly define the two branch cuts to be $[z_1, z_2]$ and $[z_3, z_4]$ so that one is inside and the other one is outside the contour. The whole integral can be written as
\begin{equation}
	W_{\bar{\rho}_\infty}(A) = \frac{\ii}{2\pi} \oint_{|z|=1} dz \frac{P_2(z) - \sinh J \sinh h \sqrt{P_4(z)}}{e^J \sinh h(z^2-1)} z^{|A|-1}\,.
\end{equation}
Inside the contour, the integrand does not have any pole but the branch cut $[z_1,z_2]$. By the Cauchy's integral theorem, we can deform the contour to enclose only the branch cut $[z_1,z_2]$. Along the deformed contour, $P_2(z)$ is analytical and thus can be ignored. We have $\arg(z-z_1) =0$, $\arg(z-z_2)= \arg(z-z_3) = \arg(z-z_4)=\pi$ above the branch cut and $\arg(z-z_1) =0$, $\arg(z-z_2)= \arg(z-z_3) = \arg(z-z_4)=-\pi$ below the branch cut. Consequently, we can convert the contour integral to the following ordinary integral
\begin{equation}
	W_{\bar{\rho}_\infty}(A) = \frac{\sinh J}{\pi e^{J}} \int_{z_1}^{z_2} dx\, \frac{\sqrt{(x-z_1)(z_2-x)(z_3-x)(z_4-x)}}{1-x^2} x^{|A|-1}\,.
\end{equation}
When the system is deep in the volume-law phase, $J\gg 1\gg h$, $z_1\ll z_2\ll 1$ and the $(z_3-x)(z_4-x)$ and $1-x^2$ factors are both of order $\scO(1)$ during the whole integral. Therefore, we approximate them by a constant, which can be fixed by comparing with the exact result, namely
\begin{equation}
	W_{\bar{\rho}_\infty}(A) \approx \frac{\sinh J}{\pi e^{J}} C \int_{z_1}^{z_2} dx\, \sqrt{(x-z_1)(z_2-x)} x^{|A|-1}\,.
\end{equation}
The rest of integral is the hypergeometric function. Recalling the Euler's formula, we have
\begin{equation}
	\label{eqn:Wsigma contour}
	\begin{aligned}
	W_{\bar{\rho}_\infty}(A) = \frac{\sinh J\, z_1^{|A|}z_2}{4e^J |A|(|A|+1)} & C \left[\left(1+\frac{z_1}{z_2} \right) F\left(-\frac{1}{2},1-|A|,1,1-\frac{z_2}{z_1} \right)  \right. \\
	& \left. - \left((1+2|A|)+(1-2|A|)\frac{z_2}{z_1} \right) F\left(\frac{1}{2},1-|A|,1,1-\frac{z_2}{z_1} \right) \right]
	\end{aligned}\,.
\end{equation}
This provides a good approximation to the exact result, as is shown in \figref{fig:entropy contour}.
To compute the large $|A|$ expansion of the above expression, we need the asymptotic expansion of the hypergeometric function $F(\alpha,\beta;\gamma;z)$ with respect to a large positive $\beta$. Let us recall the definition of (Gaussian) Hypergeometric functions
\begin{equation}
	F(\alpha,\beta;\gamma;z) = \sum_{n=0}^{\infty} \frac{(\alpha)_n(\beta)_n}{(\gamma)_n} \frac{z^n}{n!}\,,\quad (a)_n = \frac{\Gamma(a+n)}{\Gamma(a)}\,.
\end{equation}
If we assume the expansion is interchangable with the infinite sum, then $(\beta)_n$ can be replaced with its large $\beta$ expansion which leads to
\begin{equation}
	F(\alpha,\beta;\gamma;z) \sim \sum_{n=0}^{\infty} \frac{(\alpha)_n}{(\gamma)_n} \frac{(\beta z)^n}{n!} = M(\alpha,\gamma;\beta z),
\end{equation}
where $M(a,b;z)$ is the confluent hypergeometric function. As a result, the large parameter expansion of $F(\alpha,\beta;\gamma;z)$ can be reduced to the large argument expansion of $M(a,b;z)$ and we have
\begin{equation}
	\label{eqn:Wsigma asymptotics}
	W_{\bar{\rho}_\infty}(A) \approx \frac{\sinh J\, z_2}{2e^J \sqrt{\pi}}\, C\, \left(\frac{z_2}{z_1} - 1 \right)^{1/2} z_1^{|A|}e^{(1-z1/z2)|A|} |A|^{-3/2}\,,
\end{equation}
The behavior of \eqnref{eqn:Wsigma asymptotics} is also plotted in \figref{fig:entropy contour}(a) as a comparison. The discrepancy only comes from the inaccurate exponential factor in \eqnref{eqn:Wsigma asymptotics} while the power-law factor turns out to be true as verified in \figref{fig:entropy contour}(b) as well as in the main text.

\end{document}